\documentclass[11pt,a4paper]{article}
\pdfoutput=1

\usepackage{jheppub}
\usepackage{latexsym}
\usepackage{revsymb}
\usepackage{multirow}
\usepackage{color}
\usepackage[usenames,dvipsnames,svgnames,table]{xcolor}

\usepackage{graphicx}
\usepackage{epsfig}  
\usepackage{epsf}    
\usepackage{dcolumn}
\usepackage{bm}
\usepackage{dcolumn}
\usepackage{textcomp}
\usepackage{float}
\usepackage{hypcap}
\usepackage[]{hyperref}
\usepackage{adjustbox}
\usepackage{multirow}
\usepackage[utf8]{inputenc}
\usepackage[T1]{fontenc} 
\usepackage{subfigure}
\usepackage{alphalph}
\usepackage{morefloats}
\usepackage{textcomp}
\usepackage{comment}

\preprint{FERMILAB-PUB-21-260-T, \,\,IPPP/20/107} 

\title{Impact of Improved Energy Resolution on DUNE sensitivity to Neutrino Non-Standard Interactions}

\author[a]{Sabya Sachi Chatterjee,}
\affiliation[a]{Institute for Particle Physics Phenomenology, Department of Physics, Durham University, Durham, DH1 3LE, UK}
\emailAdd{sabya.s.chatterjee@durham.ac.uk}

\author[b]{P. S. Bhupal Dev,}
\affiliation[b]{Department of Physics and McDonnell Center for the Space Sciences, Washington University, St.\,Louis, MO 63130, USA}
\emailAdd{bdev@wustl.edu}

\author[c]{Pedro A. N. Machado}
\affiliation[c]{Theoretical Physics Department, Fermi National Accelerator Laboratory, Batavia, IL 60510, USA}
\emailAdd{pmachado@fnal.gov}

\abstract{
The full physics potential of the next-generation Deep Underground Neutrino Experiment (DUNE) is still being explored. 
In particular, there have been some recent studies on the possibility of improving DUNE's neutrino energy reconstruction.
The main motivation is that a better determination of the neutrino energy in an event-by-event basis will translate into an improved measurement of the Dirac $CP$ phase and other neutrino oscillation parameters.
To further motivate studies and improvements on the neutrino energy reconstruction, we evaluate the impact of energy resolution at DUNE on an illustrative new physics scenario, viz. non-standard  interactions (NSI) of neutrinos with matter.
We show that a better energy resolution in comparison to the ones given in the DUNE  conceptual and technical design reports  may significantly enhance the experimental sensitivity to NSI, particularly when degeneracies are present. 
While a better reconstruction of the first oscillation peak helps disentangling standard $CP$ effects from those coming from NSIs, we find that the second oscillation peak also plays a nontrivial role in improving DUNE's sensitivity.
}
 

\arxivnumber{2106.04597}

\begin{document}
\maketitle

\section{Introduction} 
\label{sec:intro}
Decades of 
solar, atmospheric, accelerator and reactor 
neutrino experiments have firmly established the phenomenon of neutrino oscillations among the three flavors~\cite{Zyla:2020zbs}. 
After the recent discovery of the relatively large reactor mixing angle $\theta_{13}$, the three-neutrino oscillation paradigm 
has entered a new precision era, where the known oscillation parameters are being measured with an ever-increasing accuracy. At the same time, several short, medium and long-baseline neutrino oscillation experiments, either running or in the pipeline, are poised to resolve the sub-dominant effects in oscillation data sensitive to the currently unknown oscillation parameters, namely the Dirac $CP$ phase $\delta_{\rm CP}$, the sign of the atmospheric neutrino mass-squared difference $\Delta m^2_{32}$, and the octant of the atmospheric mixing angle $\theta_{23}$.



Unraveling the neutrino properties within the three-neutrino framework has been a great success so far, when analyzing either a single experiment or the entirety of current neutrino oscillation data in terms of global fits~\cite{Capozzi:2018ubv,deSalas:2020pgw, Esteban:2020cvm}. 
Nevertheless, the quest for new physics beyond the Standard Model (SM) in the neutrino sector remains remarkably vibrant. Indeed, the phenomenon of neutrino oscillations itself suggests nonzero neutrino masses, which requires some beyond the SM (BSM) physics. Therefore, exploiting the full potential of the current and next-generation neutrino experiments to probe BSM physics is an important ongoing research topic. 

One interesting model-independent framework for parameterizing BSM physics in the neutrino sector is the so-called non-standard interactions (NSI)~\cite{Wolfenstein:1977ue}; for reviews on various aspects of NSI, see Refs.~\cite{Ohlsson:2012kf, Miranda:2015dra, Farzan:2017xzy, Dev:2019anc}.  
The NSI framework is an effective field theory below the electroweak scale that may encode the oscillation phenomenology of an entire class of new physics scenarios; see e.g. Refs~\cite{Farzan:2015doa, Farzan:2015hkd, Farzan:2016wym, Forero:2016ghr, Babu:2017olk, Dey:2018yht, Babu:2019mfe} for specific ultraviolet (UV)-complete models of NSI. The presence of NSI in either neutrino production, detection or propagation through matter can crucially affect the interpretation of the experimental data~\cite{Esteban:2019lfo, Coloma:2019mbs}. At the very least, it could serve as a foil for the three-neutrino oscillation scheme~\cite{Miranda:2004nb, Girardi:2014kca, Liao:2016hsa, Masud:2016bvp, deGouvea:2016pom, Masud:2016nuj, Agarwalla:2016fkh, Deepthi:2016erc,  Deepthi:2017gxg, Flores:2018kwk, Hyde:2018tqt, Masud:2018pig, Capozzi:2019iqn, Yasuda:2020cff, Esteban:2020itz, Agarwalla:2021twp}. Therefore, a better understanding of the NSI effects is essential for the success of the future precision neutrino experiments in answering the open questions in neutrino physics.

Since neutral-current interactions affect neutrino propagation coherently, long-baseline neutrino experiments with a well-understood beam and trajectory are an ideal place to probe matter NSI effects. Numerous studies have been performed to this effect exploring the NSI prospects in the long-baseline  experiments; see e.g. Refs.~\cite{Ota:2001pw, Blennow:2007pu, Kopp:2007ne, Blennow:2008ym, Kopp:2010qt, Adhikari:2012vc, Friedland:2012tq, Coelho:2012bp, Rahman:2015vqa, deGouvea:2015ndi,  Coloma:2015kiu, Forero:2016cmb, Huitu:2016bmb, Bakhti:2016prn, Soumya:2016enw,   Blennow:2016etl, Liao:2016bgf, Fukasawa:2016lew, Liao:2016orc,  Chatterjee:2018dyd, Han:2019zkz, Denton:2020uda, Chatterjee:2020kkm, Bakhti:2020fde, Giarnetti:2021} and references therein. In fact, a mismatch in the determination of the $\delta_{\rm CP}$ extracted from the latest data of the two running long-baseline neutrino experiments, NO$\nu$A~\cite{NOVA_talk_nu2020} and T2K~\cite{T2K_talk_nu2020}, might already be hinting towards some possible indications of nonzero NSI~\cite{Denton:2020uda, Chatterjee:2020kkm}.  

The next-generation long-baseline  experiment DUNE~\cite{Acciarri:2015uup, Abi:2020evt}, with its high-intensity neutrino beam, huge statistics, wide-band spectrum and improved systematic uncertainties, is in an excellent position to probe matter NSI~\cite{deGouvea:2015ndi, Coloma:2015kiu, Blennow:2016etl, Liao:2016orc} and BSM physics involving neutrinos in general~\cite{Abi:2020kei}. At present, the full physics potential of liquid argon time projection chamber (LArTPC) detectors is still being explored. 
Several developments and novel proposals have been put forward recently  to bolster the  capabilities of these detectors, such as sub-MeV ionization energy detection in ArgoNeuT~\cite{Acciarri:2018myr} and other LArTPC detectors~\cite{Castiglioni:2020tsu}, sub-GeV study of atmospheric neutrinos~\cite{Kelly:2019itm}, and novel $\nu_\tau$ detection strategies~\cite{Conrad:2010mh, deGouvea:2019ozk, Machado:2020yxl} at DUNE. These studies are aimed at further enhancing the DUNE sensitivity to BSM physics.  

Of particular interest for the present paper is the possibility of having an improved reconstruction of the neutrino energy by collecting all ionization energy and/or identifying each particle in an event-by-event basis~\cite{DeRomeri:2016qwo, Friedland:2018vry}.
These studies show that the current energy resolution as reported in the original DUNE Conceptual Design Report (CDR)~\cite{Acciarri:2015uup} and the DUNE Technical Design Report (TDR)~\cite{Abi:2020qib} can be considerably improved by up to a factor of 2 to 3. 
It has also been shown that the better energy resolution will contribute to a more precise determination of the Dirac $CP$ phase~\cite{DeRomeri:2016qwo}.

In this work, we will show for the first time that not only the determination of the standard oscillation parameters will benefit from an improved energy resolution, but also DUNE's sensitivity to new physics. 
We will take NSI as a general example of new physics and will analyze how the experimental sensitivity changes when going from the CDR and TDR resolutions to an improved neutrino energy resolution scenario.
For the improved case, we will take the best energy reconstruction results of Ref.~\cite{Friedland:2018vry}, which leverages particle identification and the detection of all ionization energy, including de-excitation gammas from argon nuclei.
Our findings show that the better neutrino energy resolution significantly boosts the DUNE sensitivity to NSI, improving its constraints by up to 25-30\% for several NSI parameters.
The improved energy resolution leads to a better reconstruction of both first and second oscillation peaks. 
While the first peak comprises most of the statistical power in DUNE, we find that the second peak provides a non-negligible contribution to the sensitivity, helping better disentangle standard $CP$ effects from those coming from NSIs.\footnote{ 
For discussions related to the relevance of DUNE's second oscillation peak, see e.g. Ref.~\cite{DeRomeri:2016qwo, Rout:2020emr} in the context of standard oscillation physics and Ref.~\cite{Huber:2010dx} in the context of NSI.}
These findings hold true for single NSI, with or without associated $CP$ phases, and also when multiple NSI parameters are considered at the same time.


The rest of the paper is organized as follows: In Sec.~\ref{sec:framework}, we briefly review the theoretical framework of NSIs. In Sec.~\ref{sec:setup}, we review the DUNE experimental setup. In Sec.~\ref{sec:simulation}, we give our simulation details. In Sec.~\ref{sec:resolution}, we discuss the improved energy resolution. Our main numerical results are presented in Sec.~\ref{sec:numerical-results}. In Sec.~\ref{sec:discussion}, we present a qualitative discussion of our results. Our conclusions are given in Sec.~\ref{sec:con}.


\section{Theoretical framework} 
\label{sec:framework}
Non-standard interactions, or NSIs, first recognized in Ref.~\cite{Wolfenstein:1977ue}, is an effective field theory below the weak scale, without explicit $SU(2)_L$ invariance, that encompasses four-fermion operators involving at least one neutrino field.
Logically, we can divide NSIs into two categories: charged-current (CC) operators, in which the leptonic current has a neutrino and a charged lepton; and neutral-current (NC) NSI where the leptonic current has two neutrino fields. 
While CC-NSIs may affect neutrino production and detection~\cite{Grossman:1995wx}, NC-NSIs induce non-trivial matter effects and modify the dynamics of flavor conversion in matter~\cite{Wolfenstein:1977ue}.
Long-baseline experiments like DUNE are strongly affected by matter effects, and thus are expected to have exceptional sensitivity to NC-NSIs.

While the ultraviolet completion of NC-NSIs can arise either via heavy mediators (see e.g. Refs.~\cite{Biggio:2009nt, Forero:2016ghr, Babu:2019mfe}) or light mediators~\cite{Farzan:2015doa,Farzan:2015hkd, Babu:2017olk, Datta:2017pfz}, we adopt an agnostic approach here and analyze the low energy phenomenology irrespective of model building considerations.
To be more precise, NC-NSI can be described by dimension-six operators as, 
\begin{equation}
\mathcal{L}_{\mathrm{NC\mbox{-}NSI}} \;=\;
-2\sqrt{2}G_F 
\varepsilon_{\alpha\beta}^{fC}
\bigl(\overline{\nu}_\alpha\gamma^\mu P_L \nu_\beta\bigr)
\bigl(\overline{f}\gamma_\mu P_C f\bigr)
\;,
\label{H_NC-NSI}
\end{equation}
where $\alpha, \beta = e,\mu,\tau$ denote the 
neutrino flavors,  $f = e,u,d$ indicate the matter 
fermions, $C=L, R$ corresponds to the chirality of the 
fermionic $f$-$f$ current, and $\varepsilon_{\alpha\beta}^{fC}$ are the strengths 
of the NSI. 
The hermiticity of the Lagrangian requires
\begin{equation}
\varepsilon_{\beta\alpha}^{fC} \;=\; \left(\varepsilon_{\alpha\beta}^{fC}\right)^*
\;.
\end{equation}
It is worth mentioning that the diagonal ($\alpha = \beta$) NSIs are always real whereas the non-diagonal ($\alpha \neq \beta$) NSIs are in general complex as they always appear along with their associated $CP$-phases. 

For neutrino propagation in the Earth, the potential induced by matter is the sum of the potentials induced by electrons, protons and neutrons. Since only the vector part of the current is relevant for coherent forward scattering, it is convenient to define
\begin{equation}
\varepsilon_{\alpha\beta}
\;\equiv\; 
\sum_{f=e,u,d}
\varepsilon_{\alpha\beta}^{f}
\dfrac{N_f}{N_e}
\;\equiv\;
\sum_{f=e,u,d}
\left(
\varepsilon_{\alpha\beta}^{fL}+
\varepsilon_{\alpha\beta}^{fR}
\right)\dfrac{N_f}{N_e}
\;,
\label{epsilondef}
\end{equation}
where $N_f$ is the number density of fermion $f$.
For the crust of the Earth, which is what long baseline neutrinos travel through, we can assume neutral and isoscalar matter, implying  $N_n = N_p = N_e$, 
in which case $N_u = N_d = 3N_e$.
Therefore,
\begin{equation}
\varepsilon_{\alpha\beta}\, \simeq\,
\varepsilon_{\alpha\beta}^{e}
+3\,\varepsilon_{\alpha\beta}^{u}
+3\,\varepsilon_{\alpha\beta}^{d}
\;.
\end{equation}

The presence of NSI modifies the effective Hamiltonian of neutrino propagation 
in matter, which in the flavor basis becomes
\begin{equation}
H \;=\; 
U
\begin{bmatrix} 
0 & 0 & 0 \\ 
0 & k_{21}  & 0 \\ 
0 & 0 & k_{31} 
\end{bmatrix}
U^\dagger
+
V_{\mathrm{CC}}
\begin{bmatrix}
1 + \varepsilon_{ee}  & \varepsilon_{e\mu}      & \varepsilon_{e\tau}   \\
\varepsilon_{e\mu}^*  & \varepsilon_{\mu\mu}    & \varepsilon_{\mu\tau} \\
\varepsilon_{e\tau}^* & \varepsilon_{\mu\tau}^* & \varepsilon_{\tau\tau}
\end{bmatrix}\,,
\end{equation}
where $U$ is the PMNS matrix, which, in its standard parameterization,
depends on three mixing angles ($\theta_{12},\, \theta_{13},\, \rm{and,}\, \theta_{23}$) and one $CP$-phase ($\delta_{\rm CP}$).
The quantities $k_{21} \equiv \Delta m^2_{21}/2E$ and $k_{31} \equiv \Delta m^2_{31}/2E$ represent
the solar and atmospheric wavenumbers, where $\Delta m^2_{ij} \equiv m^2_i-m^2_j$ and $E$ is the neutrino energy.
$V_{\mathrm{CC}}$ is the CC matter potential, 
\begin{equation}
V_{\mathrm{CC}} 
\;=\; \sqrt{2}G_F N_e 
\;\simeq\; 7.6\, Y_e \times 10^{-14}
\bigg[\dfrac{\rho}{\mathrm{g/cm^3}}\bigg]\,\mathrm{eV}\,,
\label{matter-V}
\end{equation}
where $Y_e = N_e/(N_p+N_n) \simeq 0.5$ is the relative electron number density in Earth's  crust and
$\rho$ is the Earth matter density, which for DUNE we take as $\rho=2.848$~g/cm$^3$. 

For convenience we introduce the dimensionless quantity $\hat{v} = V_{\mathrm{CC}}/k_{31}$, which 
gauges the sensitivity to matter effects. 
Its absolute value
\begin{equation}
|\hat{v}| 
\;=\; \bigg|\frac{V_{\mathrm{CC}}}{k_{31}}\bigg| 
\;\simeq\; 8.8 \times 10^{-2} \bigg[\frac{E}{\mathrm{GeV}}\bigg]\;,
\label{matter-v}
\end{equation}
appears in the analytical expressions of the $\nu_\mu \to \nu_e$ appearance probability and $\nu_\mu \to \nu_\mu$ survival probability, as we will see below.

Let us now discuss the appearance and survival probabilities relevant for the long-baseline experiment DUNE. 
In the presence of NSI, one can realize that the mixing angle $\theta_{13}$,
the parameter $\hat{v}$ at DUNE and non-diagonal NSIs $|\varepsilon_{\alpha\beta}|$ are small. 
While the first two have similar size $\eta\sim 0.1$, where $\eta$ is simply a small expansion parameter, for  neutrino energies of a couple of GeV or so,~
 the latter cannot be much larger than that. 
Therefore we can perform an expansion on those parameters considering them to be roughly of the same magnitude
$\mathcal{O}(\eta)$. 
We can also define $\alpha \equiv \Delta m^2_{21}/ \Delta m^2_{31} = \pm 0.03$, which would then be
$\mathcal{O}(\eta^2)$. 
 Note that at DUNE $\hat v$ can be relatively large, particularly for the high energy tail of the spectrum, and thus the expansions that we will perform on small $\hat v$ may break down if $|\varepsilon_{\alpha\beta}|$ is not much smaller than 1.
 Nevertheless it is still useful to do the expansion in order to understand the role of energy resolution in the search for NSIs. 
Our numerical analysis is, however, exact and does not rely on any such expansions.

Let us discuss the appearance and disappearance oscillation probabilities up to the third order in $\eta$  given in Refs.~\cite{Kikuchi:2008vq, Liao:2016hsa, Liao:2016orc}.
We will write the NSI parameters in general as $\varepsilon_{\alpha\beta} = |\varepsilon_{\alpha\beta} |  e^{i\phi_{\alpha\beta}}$ for non-diagonal NSIs ($\alpha\neq\beta$) while for  the diagonal NSI couplings we simply have $\varepsilon_{\alpha\alpha}$ as a real parameter (its sign is physical). 
First, we start with the appearance probability,
\begin{align}
\label{eq:Pme}
P\left(\nu_{\mu}\rightarrow \nu_e \right) \, \simeq\, &   \,\sin^22\theta_{13}s_{23}^2 f^2 + 2 \alpha s_{13} \sin2\theta_{12} \sin2\theta_{23} f g \cos(\Delta + \delta_{\rm CP}) \nonumber\\
 &+ 8 \hat{v} s_{13}s_{23} \bigg\{ |\varepsilon_{e\mu}|\left[ s_{23}^2 f^2 \cos(\phi_{e\mu} + \delta_{\rm CP}) + c_{23}^2 f g \cos(\Delta + \delta_{\rm CP} + \phi_{e\mu}) \right] \nonumber \\
 &\qquad + |\varepsilon_{e\tau}|s_{23}c_{23} \left[  f^2 \cos(\phi_{e\tau} + \delta_{\rm CP}) -  f g \cos(\Delta + \delta_{\rm CP} + \phi_{e\tau}) \right]\bigg\}, 
\end{align}
where $\Delta \equiv  \Delta m^2_{31}L/4E$ is the atmospheric oscillating frequency,
$L$ being the baseline.
 For compactness, we have used the notation
$s_{ij} \equiv \sin \theta_{ij} $, $c_{ij} \equiv \cos \theta_{ij}$, and, 
\begin{eqnarray}
\label{eq:S}
f \equiv \frac{\sin\left\{[(1 - \hat{v}(1 + \varepsilon_{ee})]\Delta\right\}}{1-\hat{v}(1 + \varepsilon_{ee})}\,, \qquad  g \equiv \frac{\sin[\hat{v}(1 + \varepsilon_{ee})\Delta]}{\hat{v}(1 + \varepsilon_{ee})}.
\end{eqnarray}
The first line of Eq.~\eqref{eq:Pme} is the standard approximated formula if $\varepsilon_{ee}\to 0$, see e.g. Refs.~\cite{Arafune:1997hd, Freund:2001pn, Akhmedov:2004ny}, while the second and third lines are the modifications induced by NSIs.
Notice that for small oscillation phase $\Delta$, both $f$ and $g$ are linear in $\Delta$. 
The leading term in $|\varepsilon_{e\tau}|$ tends to cancel for small $\Delta$, and thus we expect this NSI parameter to not strongly affect the high energy part of DUNE's far-detector spectrum.
Therefore, the impact of $|\varepsilon_{e\tau}|$ may be relatively more pronounced in the second oscillation maximum.
The observability of the second oscillation maximum is directly related to the precision of DUNE's energy reconstruction.
We will see later that an improved energy resolution will significantly enhance the sensitivity to $|\varepsilon_{e\tau}|$, and that both first and second oscillation peaks contribute appreciably for this improvement.
This effect is not present for $|\varepsilon_{e\mu}|$, and we will confirm that with our numerical analysis.

The $\nu_\mu\to\nu_\mu$ disappearance oscillation probability is affected by a different set of NSI parameters:
%
%
%
%
\begin{align}
\label{eq:Pmm}
P\left(\nu_{\mu}\rightarrow \nu_{\mu} \right) \, \simeq\, & \, 1 - \sin^22\theta_{23} \sin^2\Delta + \alpha c_{12}^2 \sin^22\theta_{23} \Delta \sin 2\Delta - 4 s_{23}^4 s_{13}^2\frac{ \sin^2[(1-\hat{v})\Delta]}{(1-\hat{v})^2} \nonumber \\
 &- \frac{\sin^22\theta_{23}s_{13}^2}{(1-\hat{v})^2} \biggl\{\hat{v}\Delta \sin 2\Delta 
 + \sin\big[(1-\hat{v})\Delta\big] \sin\big[(1 + \hat{v})\Delta\big] \biggr\} \nonumber \\
 &- 2\hat{v} |\varepsilon_{\mu\tau}| \cos \phi_{\mu\tau}\biggl(\sin^3 2\theta_{23} \Delta \sin 2\Delta 
  + 2 \sin 2\theta_{23} \cos^2 2\theta_{23} \sin^2\Delta \biggr) \nonumber \\
  & + \left[\hat{v}\sin^2 2\theta_{23}\cos 2\theta_{23}\left(\varepsilon_{\mu\mu} - \varepsilon_{\tau\tau}\right) - \frac{\hat{v}^2}{2} \sin^4 2\theta_{23}\left(\varepsilon_{\mu\mu} - \varepsilon_{\tau\tau}\right)^2 \right]\nonumber \\
  & \qquad \times (\Delta \sin 2\Delta - 2 \sin^2\Delta) \, .
\end{align}

The first two lines describe standard oscillations, while the effect of NSIs is encoded in the last 2 lines.
The leading term on $\varepsilon_{\mu\tau}$ is also multiplied by $\cos\phi_{\mu\tau}$, see the third line in the equation above.
Thus, if one marginalizes over the phase when presenting the sensitivity to this NSI parameter, the result will typically be weak for DUNE as the leading term can always be put to zero.
In experiments where matter effects are more relevant like IceCube, $\hat v\gtrsim1$, the expansion performed here breaks down and the impact of the $\phi_{\mu\tau}$ phase becomes less pronounced, see e.g. Ref.~\cite{IceCube:2021abg}.
It is also worth mentioning that the second term in the parenthesis is proportional to $\cos^22\theta_{23}$, which may be fairly suppressed as $\theta_{23}$ approaches maximality.

The leading $\varepsilon_{\mu\mu}-\varepsilon_{\tau\tau}$ term is suppressed by $\cos^22\theta_{23}$, and thus sensitivity to this parameter should present a strong dependence on $\theta_{23}$. 
Moreover, both leading and next-to-leading terms cancel out for small $\Delta$, and so we anticipate an important role in the experimental sensitivity to this parameter to be played by the second oscillation and the energy resolution as a consequence. Our numerical simulations will confirm this expectation as well. Also, note that the first term in parenthesis in the last line of Eq.~\eqref{eq:Pmm} is linear in $\varepsilon_{\mu\mu}-\varepsilon_{\tau\tau}$, whereas the second term is quadratic. 
This is the reason for the difference in the DUNE sensitivities to $\varepsilon_{\mu\mu}$ and $\varepsilon_{\tau\tau}$, when each NSI is considered at a time, as we will see later.
In the expressions above, for normal mass ordering, the sign of $\Delta$, $\alpha$ and $\hat v$ are all positive, while for inverted ordering they are all negative. Analogous expressions to those in Eqs.~(\ref{eq:Pme}) and (\ref{eq:Pmm}) for $P\left(\nu_{\mu}\rightarrow \nu_{e} \right)$ and $P\left(\nu_{\mu}\rightarrow \nu_{\mu} \right)$ can be obtained for antineutrinos by flipping the sign of all $CP$ phases and $\hat v$. For concreteness, we will present all our numerical results for normal ordering (NO) only.


\section{Experimental setup} 
\label{sec:setup}
DUNE is a future long-baseline accelerator neutrino experiment, where neutrinos are expected to travel a distance of 1300 km from the source at Fermilab to the far-detector placed deep underground at the Sanford Lab in South Dakota. 
In order to estimate the sensitivity of the experiment to NSIs, we adopt three benchmark experimental configurations for DUNE: the Conceptual Design Report configuration (CDR)~\cite{Acciarri:2015uup,Alion:2016uaj}; the more recent Technical Design Report configuration (TDR)~\cite{Abi:2020qib,Abi:2021arg}; and a third which is the TDR configuration with improved energy resolution based on the findings of Ref.~\cite{Friedland:2018vry}, as we will discuss later, which we refer to as the Best Reconstruction (Best Rec.).

To perform the experimental simulations we have used the GLoBES package~\cite{Huber:2004ka,Huber:2007ji} together with the NSI tool from Ref.~\cite{Kopp_NSI}.
In all configurations we have assumed a 40 kton LArTPC detector.
The CDR beam configuration assumes an 80 GeV proton beam with 1.07 MW beam power resulting in $1.47\times 10^{21}$ POT/year, while the TDR uses a 120 GeV proton beam with 1.2 MW beam power resulting in $1.1\times 10^{21}$ POT/year.
We have also assumed equal time of 3.5 years running in each forward and reverse horn configurations (that is, equal neutrino and antineutrino runtime), which results in 300 and 336~kton-MW-year exposures for the CDR and TDR setups, respectively. Incidentally, the TDR staged 7-year running gives the same 336~kton-MW-year exposure~\cite{Abi:2020qib}.
For the CDR and TDR simulations, specific details on systematic errors and efficiencies can be found in Refs.~\cite{Alion:2016uaj, Abi:2021arg}.



\section{Simulation details} 
\label{sec:simulation}
In order to quantify the statistical sensitivity of all the numerical simulations performed in this work,   
we use the built-in $\chi^2$ function in GLoBES, which incorporates systematic uncertainties via pull parameters and penalty terms (see Ref.~\cite{Huber:2007ji} for details).
The total $\chi^2$ is the sum of all four channels that can be studied at DUNE: muon  disappearance and electron  appearance, in both forward and reverse horn polarity (i.e. neutrino and antineutrino modes).
We marginalize over SM and new physics parameters whenever specified to obtain the minimum $\Delta\chi^2$. 
 
In all our numerical simulations we have used a line-averaged constant Earth matter density $\rho$ = 2.848 gm/$\rm{cm}^3$ for DUNE  following the PREM~\cite{stacey:1977, PREM:1981} profile. 
Unless otherwise stated, we have used the following benchmark values of the standard three-neutrino oscillation parameters: $\Delta m_{21}^2 = 7.5\times 10^{-5}\,\rm eV^2$, $\Delta m_{31}^2 = 2.51\times 10^{-3}\,\rm eV^2$, $\sin^2\theta_{12} = 0.310$, $\sin^2\theta_{13} = 0.022$, and $\sin^2\theta_{23} = 0.56$. 
Moreover, throughout this paper for simplicity we have assumed normal mass ordering (NO).
Note that our benchmark parameters closely agree with the current global fit results~\cite{Capozzi:2018ubv, deSalas:2020pgw, Esteban:2020cvm}.
As the solar mixing parameters have very small impact on DUNE, we have fixed $\theta_{12}$ and $\Delta m_{21}^2$ to their above mentioned values.
Atmospheric parameters $\Delta m^2_{31}$, $\theta_{23}$ and $\delta_{\rm CP}$ are always allowed to vary freely as their measurements are among the main goals of DUNE. We adopt a 3.7\% uncertainty on the reactor angle, which is better constrained at experiments like Daya Bay~\cite{Adey:2018zwh} and RENO~\cite{Bak:2018ydk}. Finally, we assume a 5\% uncertainty on the average matter density, slightly larger than the 2\% uncertainty assumed in the TDR.
Whenever appropriate, we also marginalize over some new physics parameters.
\section{Improved energy resolution} 
\label{sec:resolution}
Charged particles traversing LArTPC  ionize the liquid argon, leading to free electrons which drift against the electric field and are collected by wires.
The charge deposit is used to reconstruct objects, such as tracks or showers, which are then used for particle identification.
As the amount of energy deposited depends on the particle momentum and mass, the identification, together with the overall energy deposited in the detector via ionization is used to reconstruct the energy of each individual particle.
With those energies and momenta, the incoming neutrino energy is inferred.
This is to say that the reconstruction of the incoming neutrino energy, a crucial step to understand neutrino oscillations, is not a simple business.

Recently, there has been some re-evaluation of the best resolution that could be achieved by the DUNE experiment~\cite{DeRomeri:2016qwo, Friedland:2018vry}. 
In particular, Ref.~\cite{Friedland:2018vry} studies the best case scenario, regarding incoming neutrino energy reconstruction, if the detector is able to identify all particles in an event and if the small but frequent energy deposits from recoiled neutrons are also accounted for.
The authors find that a significant enhancement, up to a factor of 4, in the reconstruction with respect to the CDR values could be achieved.~\footnote{Note that in Ref.~\cite{Friedland:2018vry}, besides the ``best reconstruction,'' another possibility is also studied for reconstructing the neutrino energy, that is, adding all calorimetric energy of the hadronic system without relying on individual particle identification. It is observed that this method leads to a neutrino energy resolution comparable to the one in DUNE's TDR, as so we do not present this ``Charge'' case separately here.}

\begin{figure}[t]
\centering
\includegraphics[height=7.3cm,width=7.3cm]{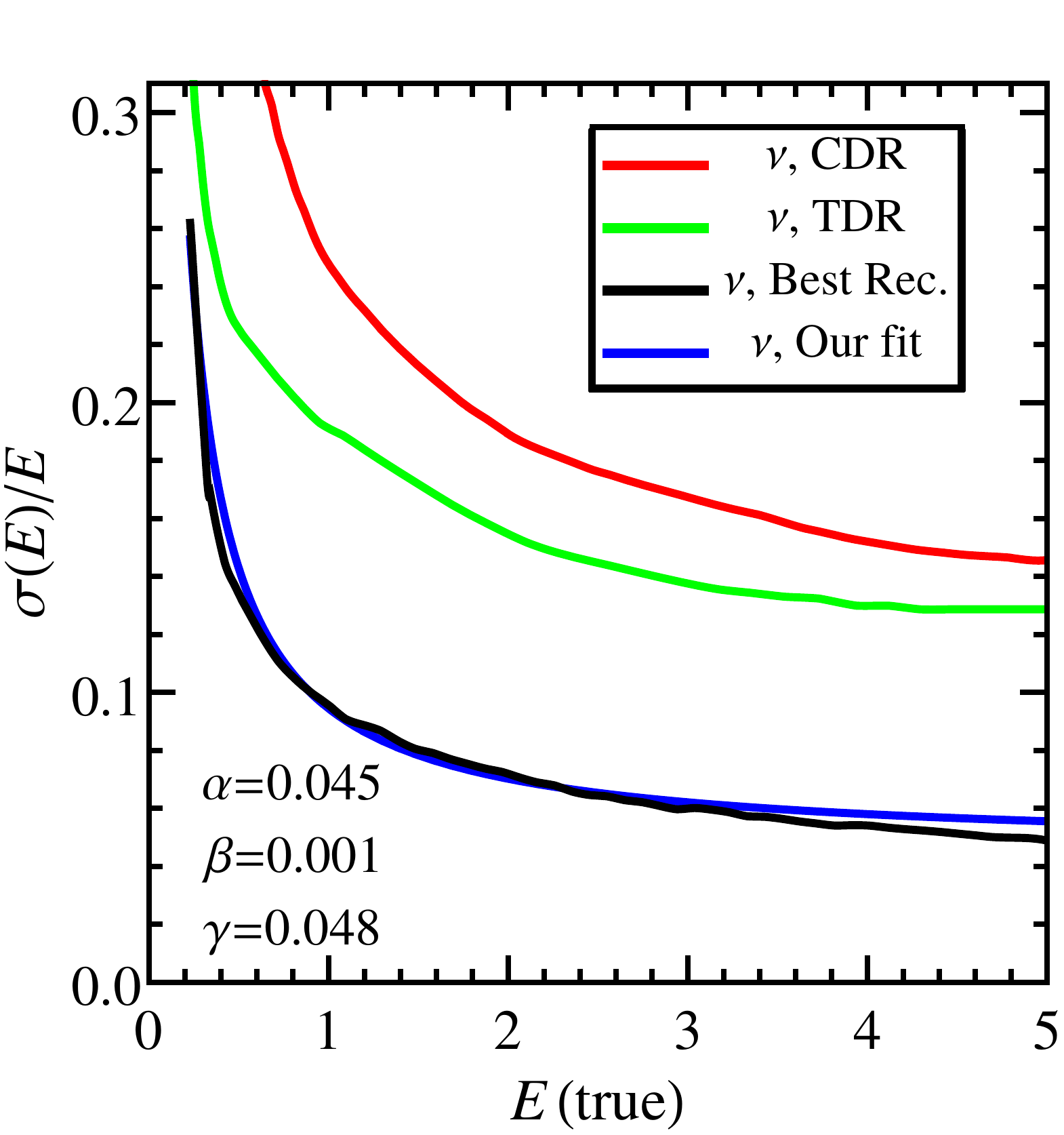}
\includegraphics[height=7.3cm,width=7.3cm]{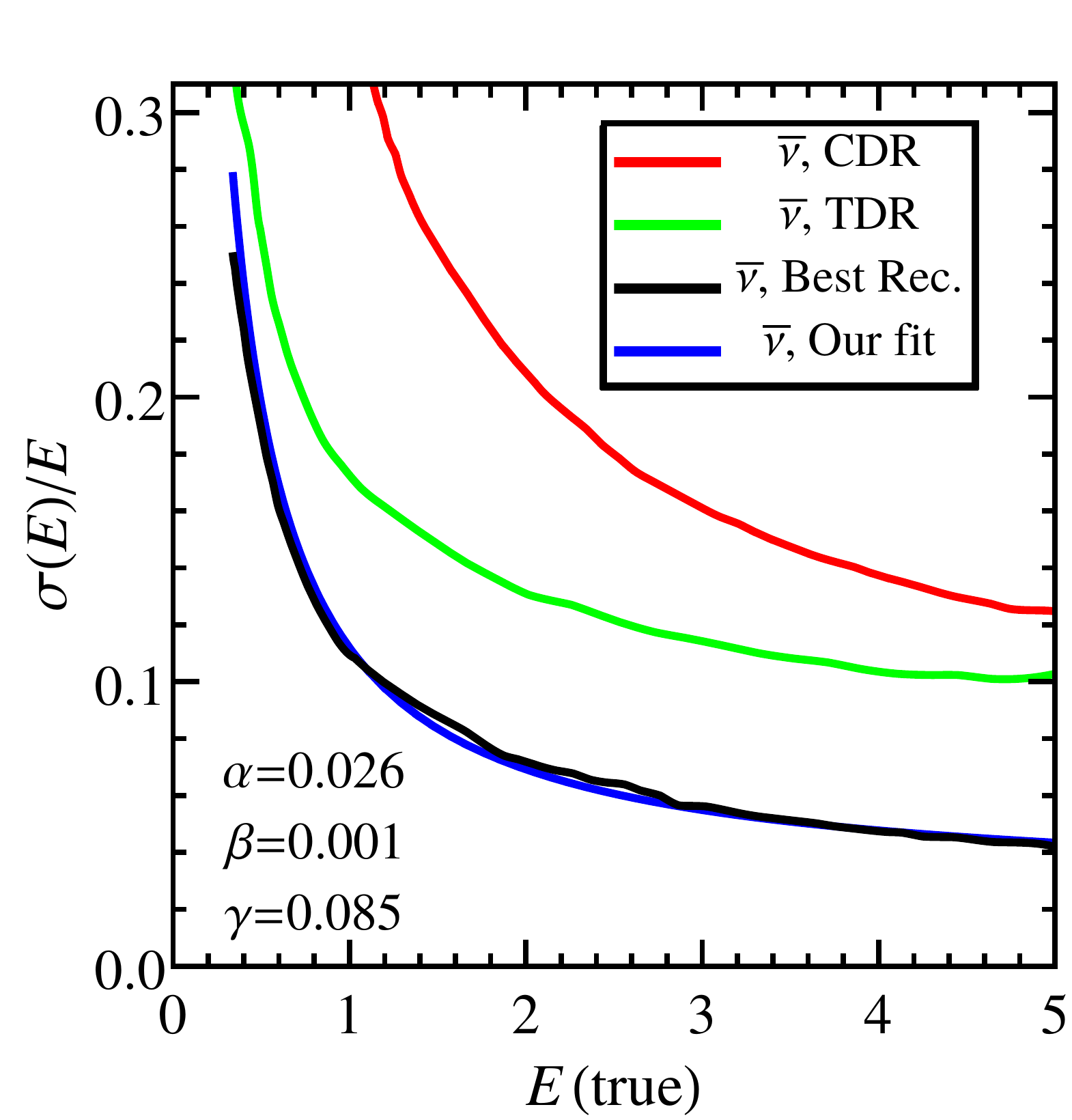}
\caption{Energy resolution as a function of true neutrino energy. Left (right) panel corresponds to neutrinos (antineutrinos).  
The red and green  curves depict the energy resolutions from DUNE's CDR and TDR, respectively, while the best reconstruction scenario from Ref.~\citep{Friedland:2018vry} is shown in black.
Our fit to the best reconstruction scenario is given by the blue curve. 
The numbers $\alpha$, $\beta$, and $\gamma$ in each panel correspond to the best fit values used in Eq.~\eqref{en:res}.  }
\label{fig:resolution}
\end{figure}

\begin{figure}[t]
\centering
\includegraphics[height=7.3cm,width=7.3cm]{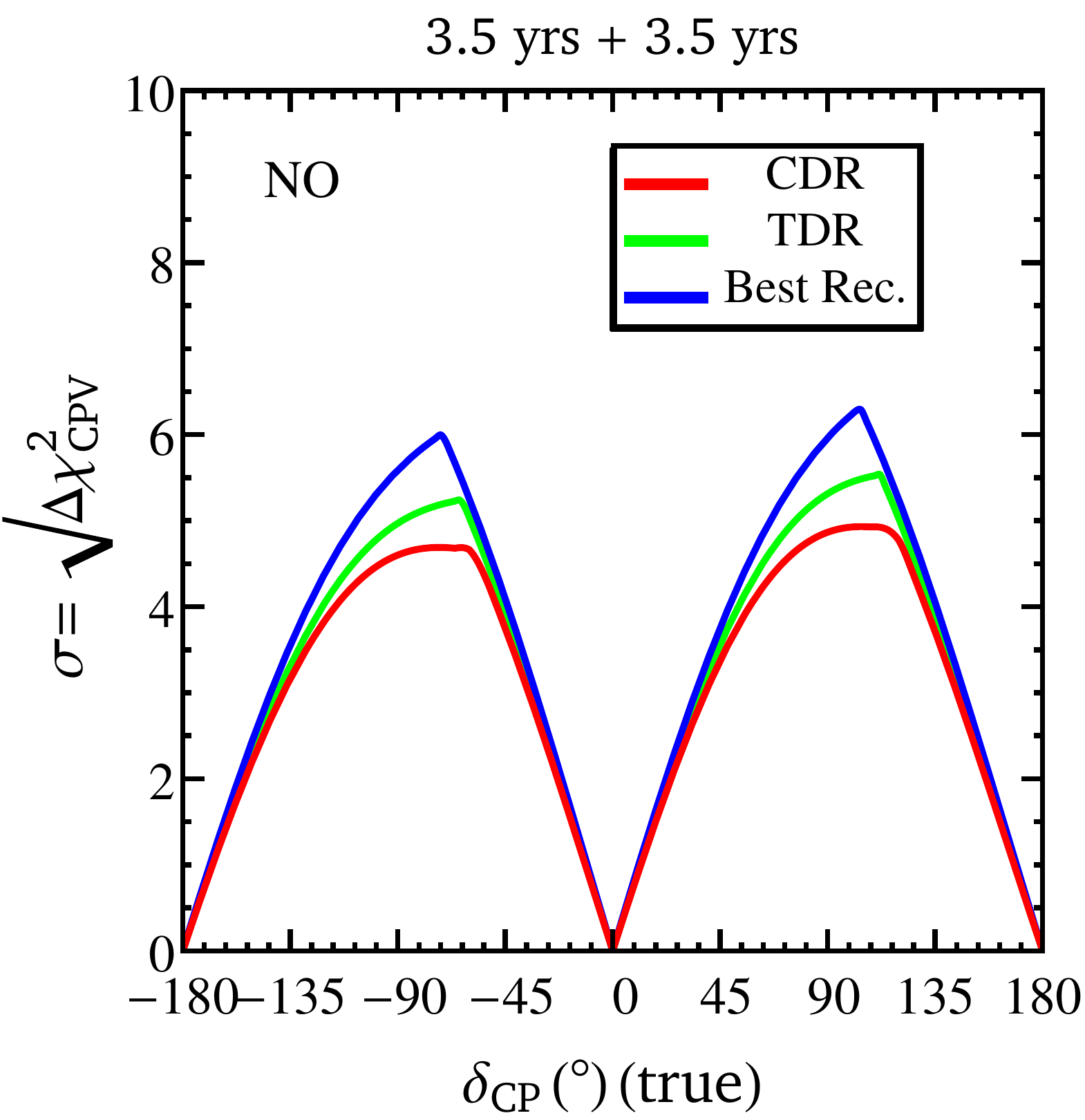}
\includegraphics[height=7.3cm,width=7.3cm]{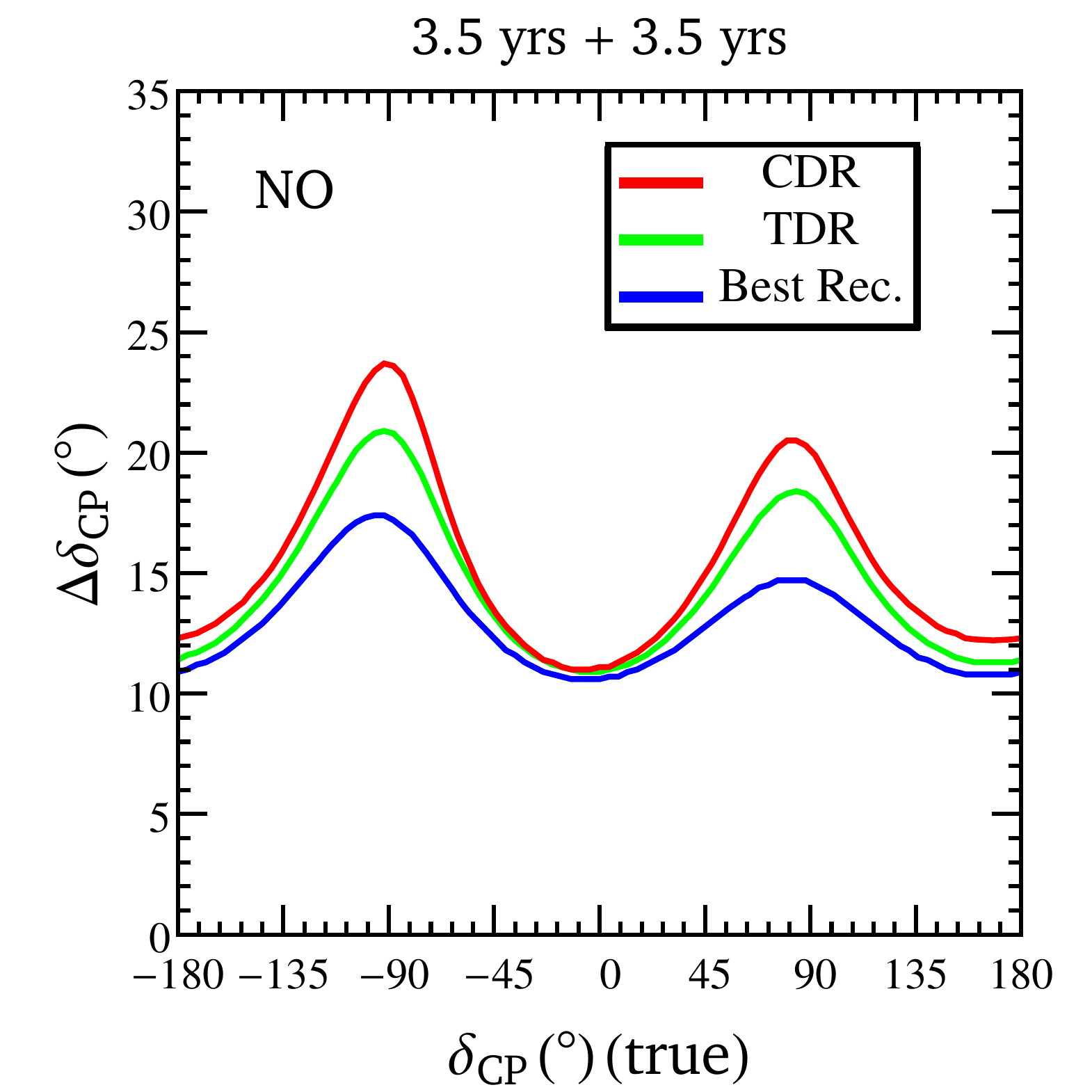}
\caption{\textbf{Left panel}: $CP$ violation discovery ($\delta_{\rm CP}(\textrm{test}) \neq 0, \pm \pi$) potential of DUNE with different assumptions on the DUNE energy resolution. \textbf{Right Panel}: 1$\sigma$ ($\Delta\chi^2 = 1$) uncertainties on the true value of $\delta_{\rm CP}$ determined by DUNE. Red, green and blue curves represent the CDR, TDR and Best Reconstruction setup respectively. We have assumed normal ordering (NO) in the analysis for both plots.}
\label{fig:delcp_precision}
\end{figure}

In this paper, we are interested in understanding how much an improved energy resolution would enhance DUNE's sensitivity to new physics, particularly NSIs. In Fig.~\ref{fig:resolution} we present the energy resolution for neutrinos (left panel) and antineutrinos (right panel) as a function of the true neutrino energy for the CDR (red line, worst resolution), and for the TDR (green line).
We also show the best case scenario for the energy resolution obtained in Ref.~\cite{Friedland:2018vry} (black line, labeled ``Best Rec.'') as well as a simple parameterization we perform to reproduce the best case results (blue line, labeled ``Our fit''). Note that 
from Fig.~\ref{fig:delcp_precision} onwards we use our parameterization of the energy resolution in the Best Reconstruction case.

Our parameterization of the Best Reconstruction case is obtained by using an energy resolution function $R(E,E_r)=e^{-(E-E_r)^2/2\sigma^2}/\sigma\sqrt{2\pi}$, where $E$ is the true neutrino energy, $E_r$ is the reconstructed energy, and the energy resolution $\sigma$ is given by
\begin{eqnarray}
\label{en:res}
\sigma(E)/{\rm GeV} = \alpha.(E/{\rm GeV}) + \beta.\sqrt{E/ {\rm GeV}} + \gamma \, ,
\end{eqnarray}
where, $(\alpha,\,\beta,\,\gamma)$ are the parameters for the fit: $(0.045,\, 0.001,\, 0.048)$ for neutrinos and $(0.026,\,0.001,\,0.085)$ for antineutrinos (cf. Fig.~\ref{fig:resolution}). 
We assume the same energy resolution for the appearance and disappearance modes.
To simulate the neutral current, $\nu_e$ contamination, misidentified muon, and $\nu_\mu\to\nu_\tau$ backgrounds we have used the same energy resolution migration matrices as provided in Refs.~\cite{Alion:2016uaj} (CDR) and \cite{Abi:2021arg} (TDR and Best Reconstruction).

As we can see from Fig.~\ref{fig:resolution}, the energy resolution is the worst for the CDR configuration, somewhat better for the TDR and improves considerably in the Best Reconstruction case. 
Quantitatively, for neutrinos at 1 GeV we have about 25\%, 19\% and 9\% energy resolution, respectively.
For reference, near 0.9~GeV, an energy resolution of about 20\% would completely wash out the second oscillation maximum.

A concrete example illustrating the importance of the energy resolution for DUNE is presented in Fig,~\ref{fig:delcp_precision}, where we show, as a function of the true $\delta_{\rm CP}$ phase, the $CP$ violation sensitivity (left panel)~\footnote{The $CP$ violation sensitivity is defined as the statistical significance at which one can reject the test hypothesis of no $CP$ violation i.e., $\delta_{\rm CP}(\rm{test}) = 0, \rm{or}\pm \pi$.} and the precision on $\delta_{\rm CP}$ achievable at DUNE at 1$\sigma$ ($\Delta\chi^2=1$, right panel) for 3.5 years in each neutrino and antineutrino mode.
For both panels, the red, the green, and the blue lines correspond to the CDR, TDR, and the Best Reconstruction benchmarks. 
As discussed above we marginalize over all relevant oscillation parameters except the solar ones, as well as over the matter density.
As we can see, the effect of a better energy resolution is non-negligible, substantially enhancing the $CP$ precision, in particular for maximal $CP$ violation.
Quantitatively, for $\delta_{\rm CP}(\rm{true}) = -90^{\circ}$, the $CP$ precision is $\sim 24^\circ$ for the CDR, $\sim 21^\circ$ for the TDR, and $\sim 17^\circ$ for the Best Reconstruction case.
Similar findings have been reported previously in Ref.~\cite{DeRomeri:2016qwo}. 

It is important to mention that the $CP$ precision obtained here with DUNE's GLoBES simulation files~\cite{Abi:2021arg} does not match the official DUNE version of the plot given in Ref.~\cite{Abi:2020qib}. 
The main reason seems to be the treatment of systematics related to the flux and/or cross section shape uncertainties.
Compared to a full-fledged Monte Carlo simulation, the GLoBES treatment of spectral uncertainties in Ref.~\cite{Abi:2021arg} is somewhat simplified.

\section{Numerical Results}
\label{sec:numerical-results}
In this section we discuss the main results we have obtained from our numerical simulations regarding the impact of the energy resolution on the sensitivity to NSIs. 

\begin{figure}[t!]
\centering
\includegraphics[height=4.9cm,width=4.9cm]{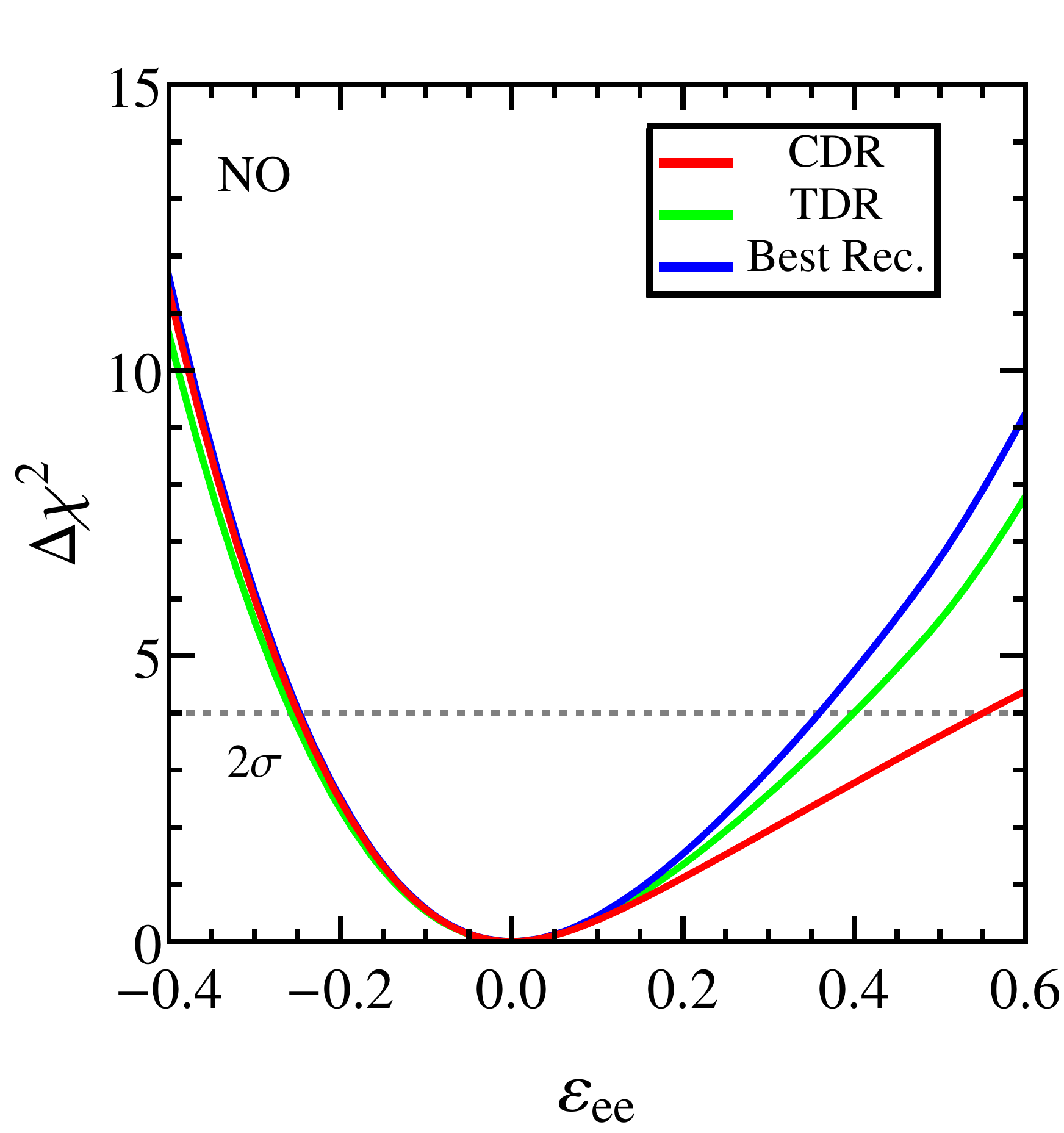}
\includegraphics[height=4.9cm,width=4.9cm]{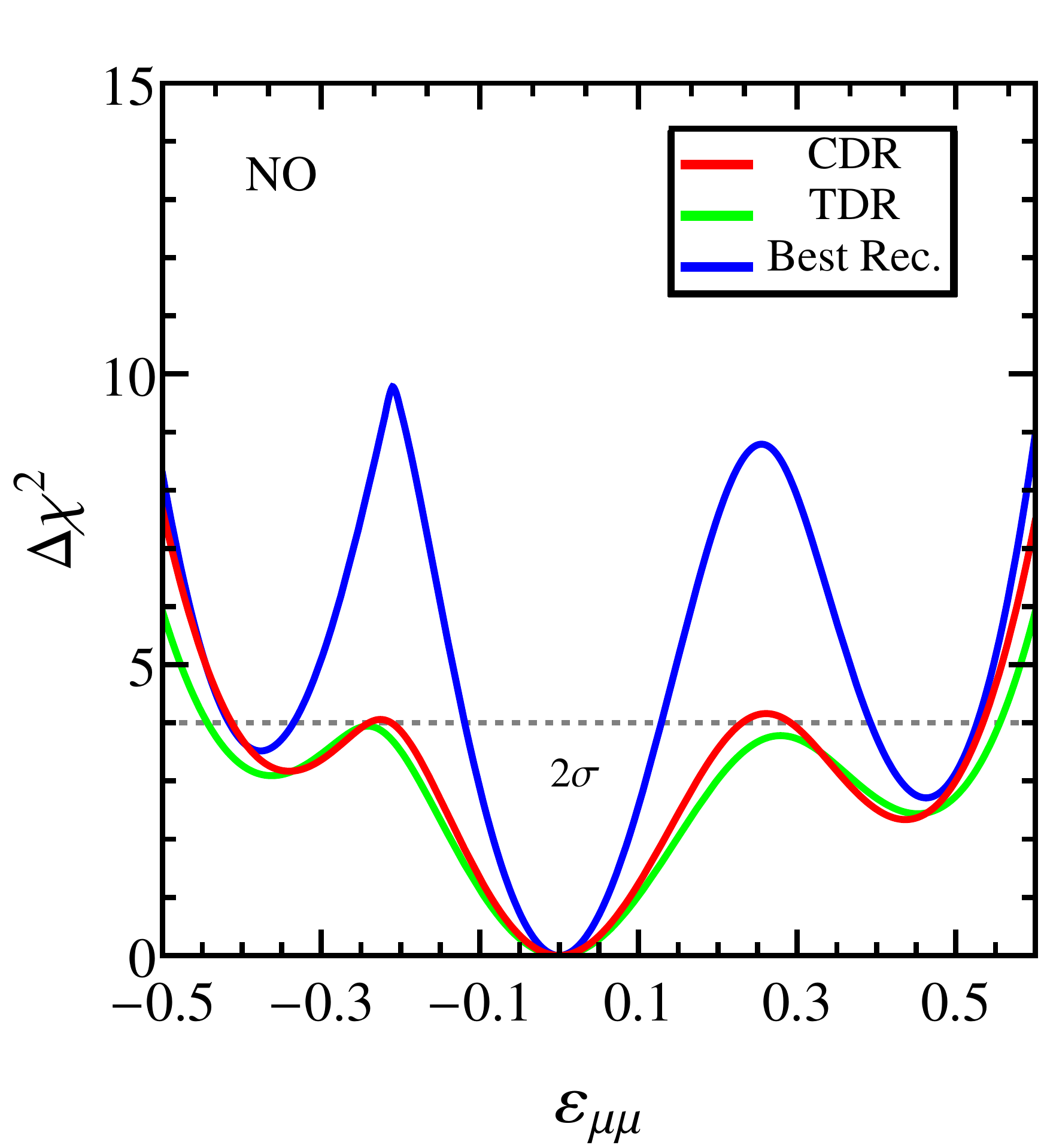}
\includegraphics[height=4.9cm,width=4.9cm]{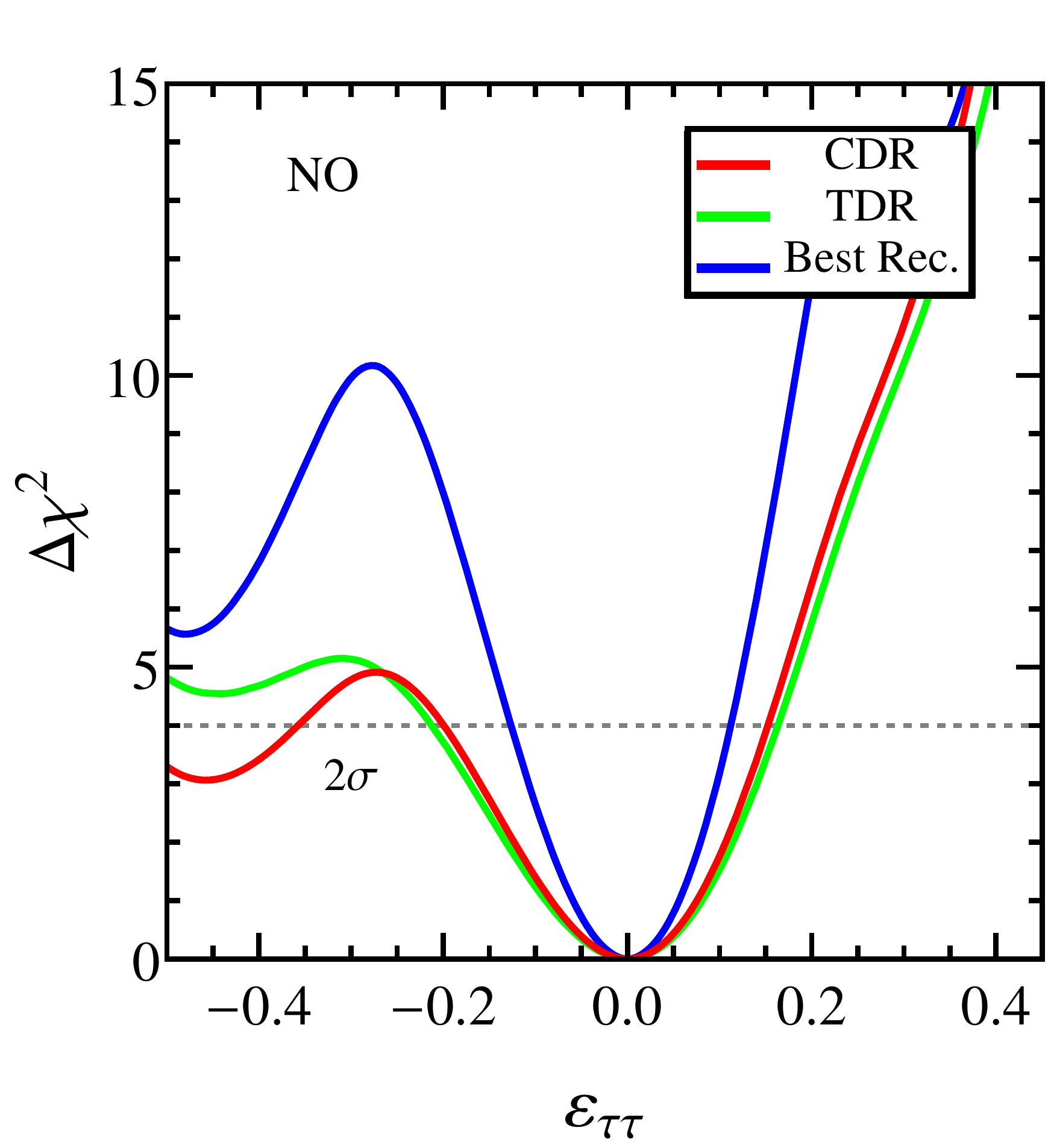}\\
\includegraphics[height=4.9cm,width=4.9cm]{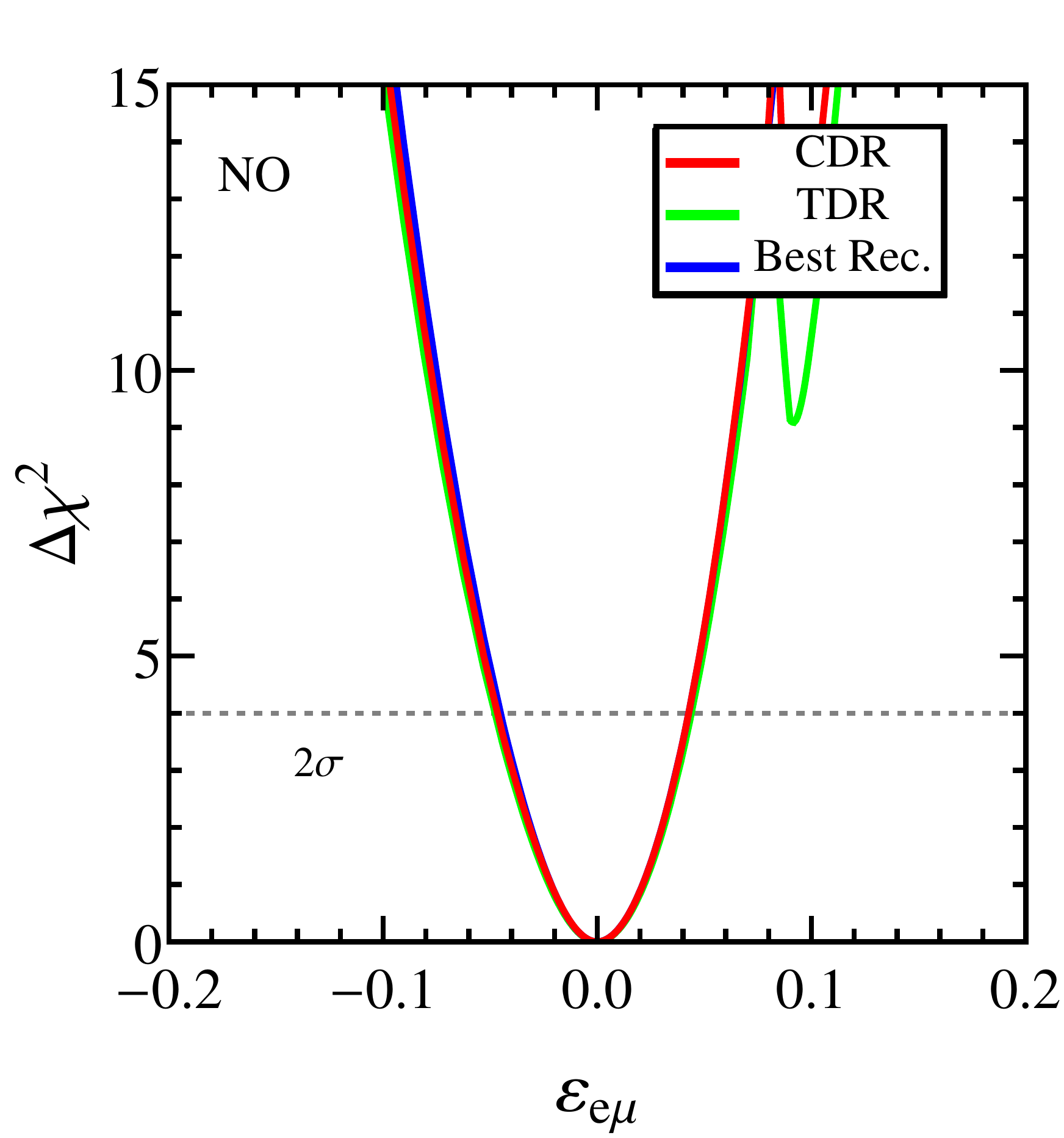}
\includegraphics[height=4.9cm,width=4.9cm]{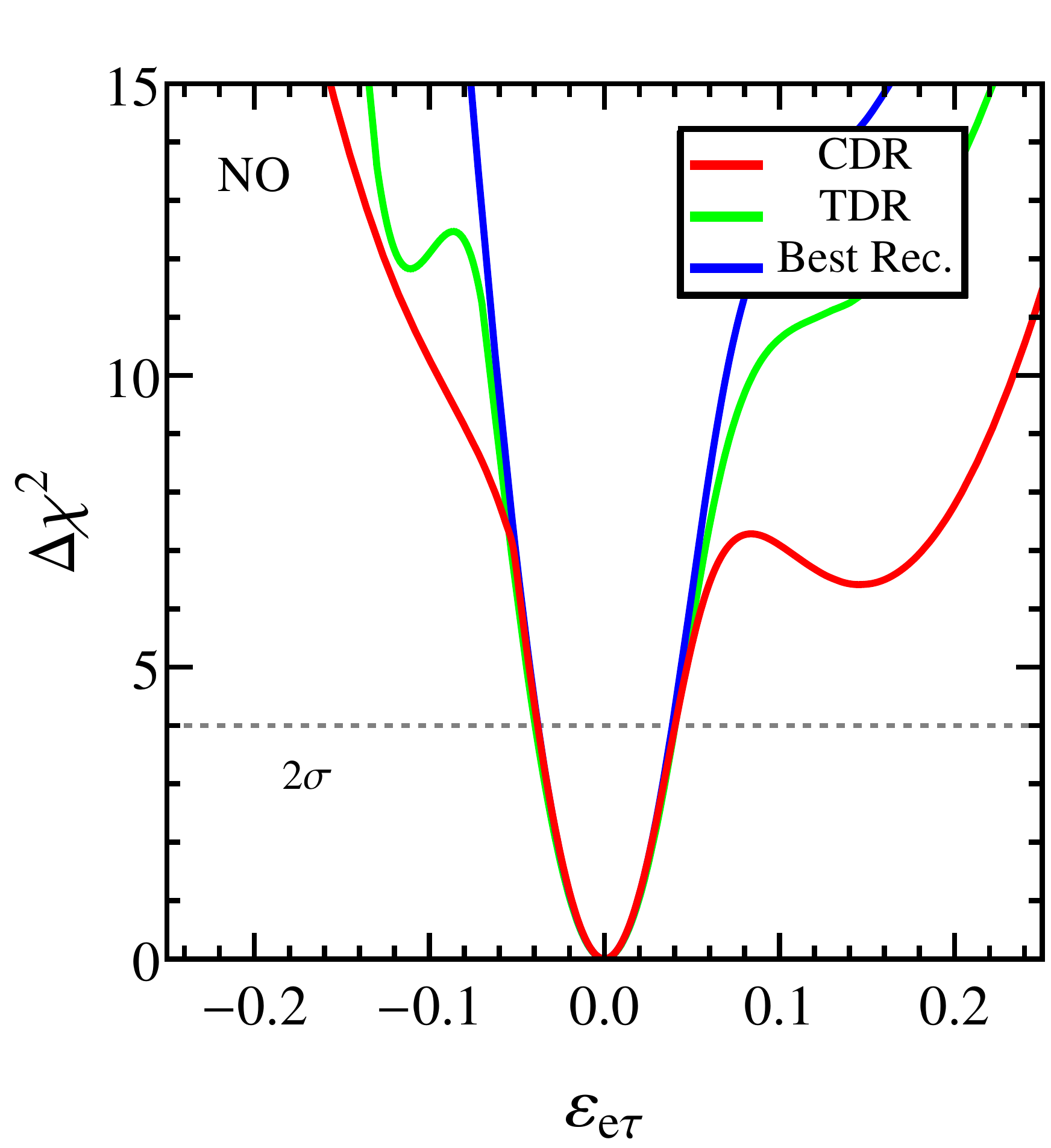}
\includegraphics[height=4.9cm,width=4.9cm]{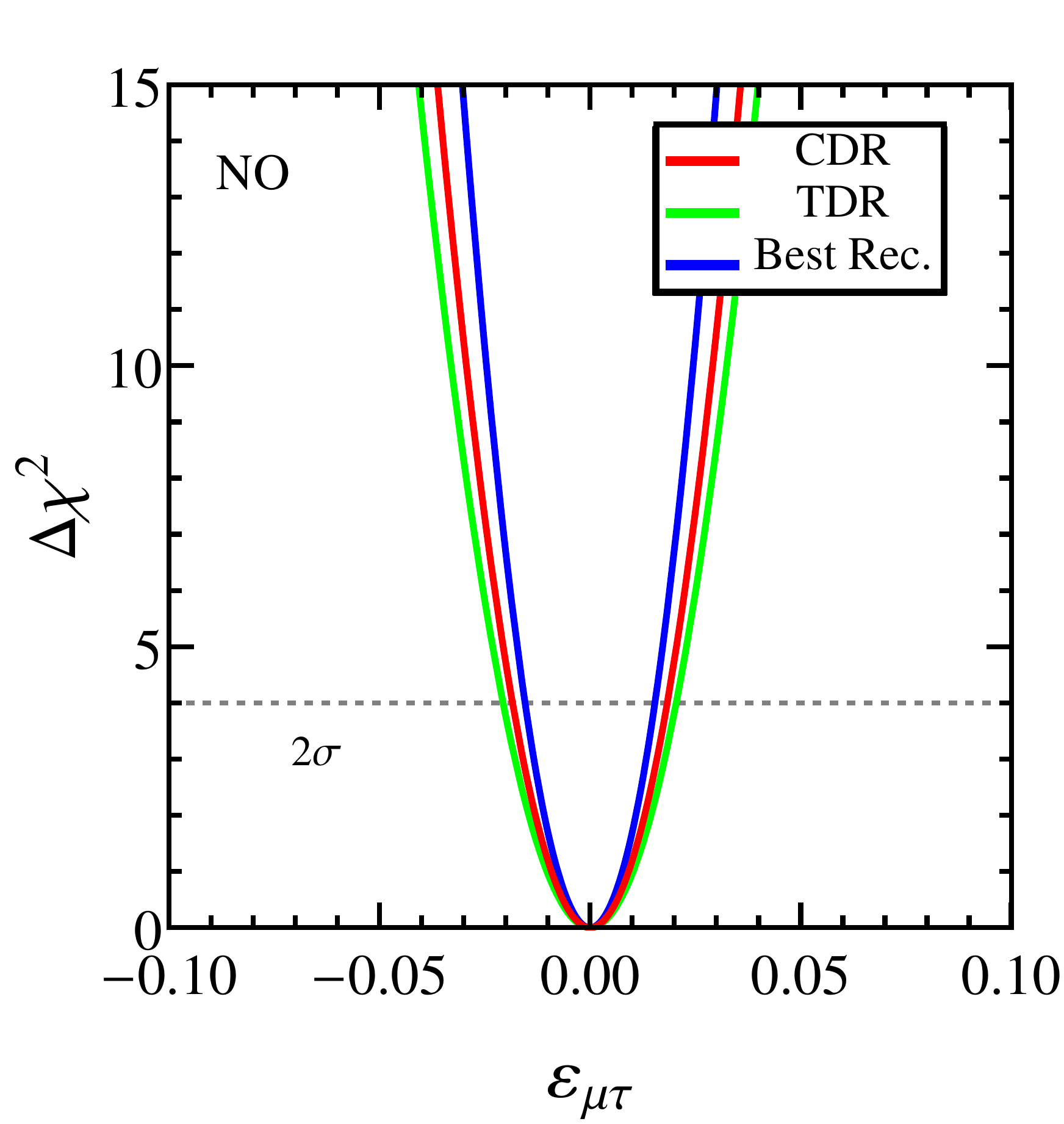}
\caption{One-dimensional projections of the DUNE sensitivity to the diagonal NSI parameters $\varepsilon_{ee},\varepsilon_{\mu\mu},\varepsilon_{\tau\tau}$ in the upper panel and non-diagonal $\varepsilon_{e\mu},\varepsilon_{e\tau},\varepsilon_{\mu\tau}$ in the lower panel. 
The red line represents the result with ``CDR'' configuration, black line is for the ``TDR'' configuration, and the blue line corresponds to the ``Best Reconstruction'' configuration. Normal mass ordering is assumed in the data as well as in the theory. We have considered $\delta_{\rm CP}(\rm{true}) = -90^{\circ}$ and the non-diagonal NSI parameters as real.}
\label{chisq_proj}
\end{figure}

Let us start with Fig.~\ref{chisq_proj}, where we have adopted a simplified framework in which all NSI parameters $\varepsilon_{\alpha\beta}$ are taken to be real parameters and we take one at a time when analyzing DUNE's sensitivity.
Here, and throughout this paper, we will always assume standard three neutrino oscillations as the true hypothesis and test nonzero NSI against it, that is our $\Delta\chi^2 = \chi_{\rm NSI}^2 - \chi_{\rm SM}^2$.
For concreteness we have assumed $\delta_{\rm CP}(\textrm{true})\,=\,-90^{\circ}$, which is close to the current preferred value from global fits~\cite{Capozzi:2018ubv, deSalas:2020pgw, Esteban:2020cvm}. 
We marginalize on the two mixing angles $\theta_{13}$ and $\theta_{23}$, the standard $CP$ phase $\delta_{\rm CP}$, the mass-squared splitting $\Delta m^2_{31}$, and the matter density, as discussed in the previous section.
The CDR, TDR and Best Reconstruction scenarios are shown in red, green and blue, respectively.
The upper panels show the sensitivity for the diagonal NSI parameters whereas the lower panel shows the sensitivity for the non-diagonal NSI parameters, which are assumed real here. 

\begin{table*}[t!]
{%
\newcommand{\mc}[3]{\multicolumn{#1}{#2}{#3}}
\newcommand{\mr}[3]{\multirow{#1}{#2}{#3}}
\begin{center}
\begin{tabular}{|c|c|c|c|}\hline
 NSI  Parameter& CDR & TDR & Best Rec. \\
\hline\hline
\mr{2}{*}{$\varepsilon_{ee}$} &\mr{2}{*}{$[-0.249 , \,  +0.552]$} & \mr{2}{*}{$[-0.256, \, +0.399]$}  & \mr{2}{*}{$[-0.246 , \, +0.360]$} \\
 &   &  & \\
 \hline
\mr{4}{*}{$\varepsilon_{\mu\mu}$} &\mr{2}{*}{$[-0.415,\, -0.240]$,} & \mr{4}{*}{$[-0.445 , \, +0.549]$}  & \mr{2}{*}{$[-0.416 ,\, -0.335],$ } \\
& \mr{2}{*}{$[-0.214,\,0.232]$,} & & \mr{2}{*}{$[-0.117 ,\, 0.128],\,$ } \\
 & \mr{2}{*}{$[0.289,\, 0.522]$}  &  & \mr{2}{*}{$[0.393,\, 0.520]$} \\
  &   &  &  \\
 \hline
 \mr{3}{*}{$\varepsilon_{\tau\tau}$} & \mr{2}{*}{$[-0.550 , \, -0.357]$, } & \mr{3}{*}{$[-0.214 , \, +0.164]$}  & \mr{3}{*}{$[-0.126 , \, +0.112]$} \\
 & \mr{2}{*}{$[-0.200 , \, +0.154]$ } & & \\
 & & & \\
 \hline
 \mr{2}{*}{$\varepsilon_{e\mu}$}&\mr{2}{*}{$[-0.046, \, +0.043]$} & \mr{2}{*}{$[-0.047 , \, +0.045]$} & \mr{2}{*}{$[-0.045 , \,  +0.043] $} \\
 &  &  & \\
\hline
\mr{2}{*}{$\varepsilon_{e\tau}$} &\mr{2}{*}{$[-0.038 , \,  +0.041]$} & \mr{2}{*}{$[-0.039 , \, +0.041]$}  & \mr{2}{*}{$[-0.033 , \, +0.033] $} \\
 &  &  & \\
\hline
\mr{2}{*}{$\varepsilon_{\mu\tau}$} &\mr{2}{*}{$[-0.018 , \, +0.018]$} & \mr{2}{*}{$[-0.021 , \, +0.021]$}  & \mr{2}{*}{$[-0.015 , \, +0.015] $} \\
 &   &  & \\
 \hline
\end{tabular}
\end{center}
}%
\caption{DUNE sensitivities on the NSI parameters at $2\sigma$ confidence level (C.L.) with 1 degree of freedom (d.o.f.) (i.e.~$\Delta\chi^2=4$), assuming  normal mass ordering
and $\delta_{\rm CP}(\rm true)=-90^\circ$ (cf. Fig.~\ref{chisq_proj}). 
Left, middle, and right columns correspond to CDR, TDR,  and Best Reconstruction cases respectively. Here the non-diagonal NSI parameters have been assumed real. }
\label{table1}
\end{table*}

We can clearly see that, while the CDR and TDR cases yield comparable results, an improved energy reconstruction significantly enhances DUNE's sensitivity to NSIs, in particular for $\varepsilon_{\mu\mu}$, $\varepsilon_{\tau\tau}$, and $\varepsilon_{e\tau}$, as well as $\varepsilon_{\mu\tau}$ to a certain extent.
There is some impact in $\varepsilon_{ee}$, though somewhat small, and almost no effect on
$\varepsilon_{e\mu}$. 
Numbers comparing the allowed regions at $2\sigma$ ($\Delta \chi^2=4$) for all NSI parameters and for the three cases can be found in Table~\ref{table1}.
Note that there are existing constraints from several experiments on all NSI parameters, see e.g. Ref.~\cite{Farzan:2017xzy}.

Until now we have treated the non-diagonal NSI parameters as real parameters. 
However it might be interesting to see how the bound changes when we treat these parameters as complex. 
As we discussed before, the NSI $CP$ phases may significantly impact the sensitivity to $\varepsilon_{\mu\tau}$, as the first order contribution of this NSI parameter to the $\nu_\mu$ disappearance probability goes like $|\varepsilon_{\mu\tau}|\cos\phi_{\mu\tau}$, see Eq.~\eqref{eq:Pmm}.

\begin{figure}[t!]
\centering
\includegraphics[height=4.9cm,width=4.9cm]{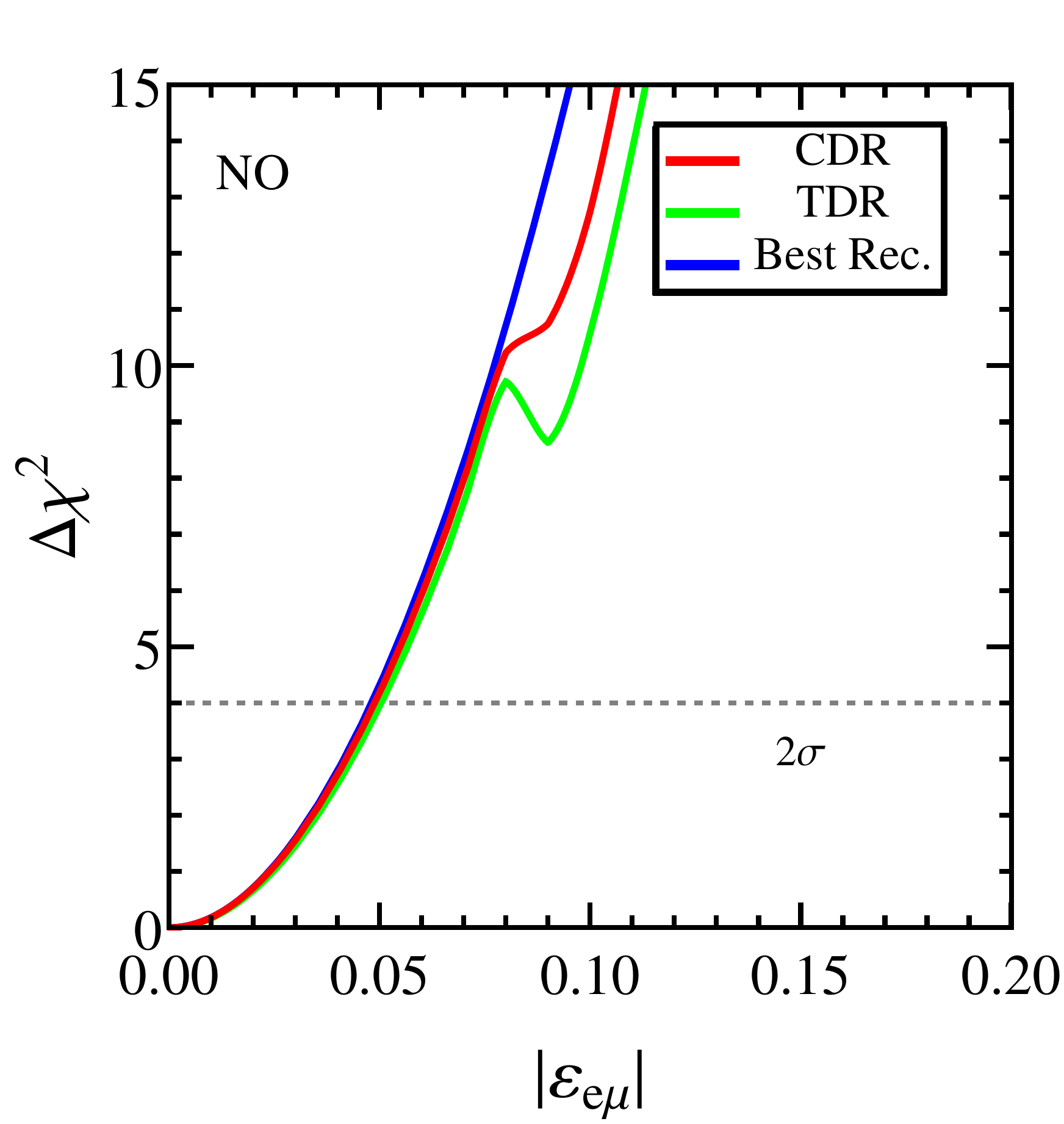}
\includegraphics[height=4.9cm,width=4.9cm]{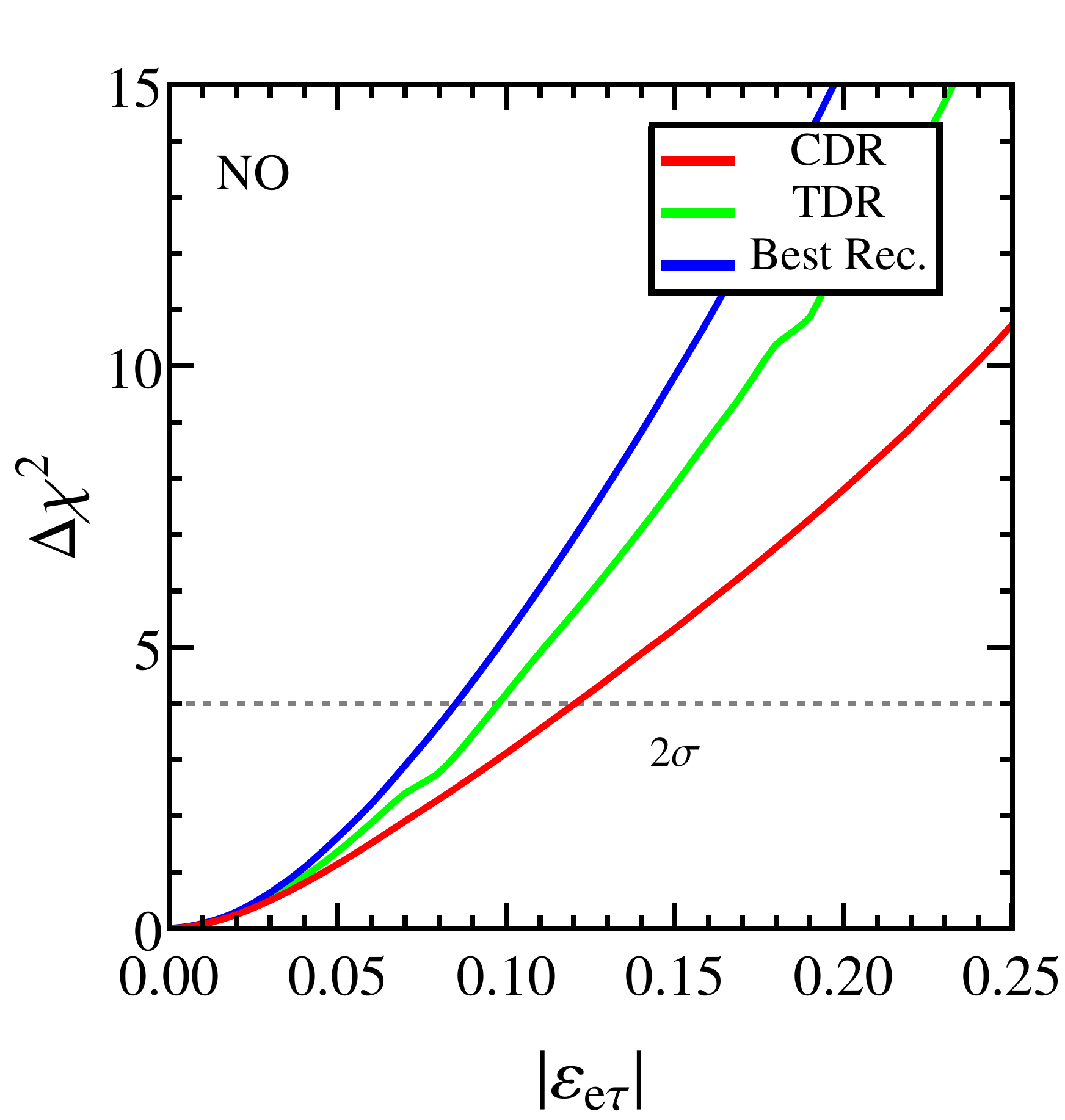}
\includegraphics[height=4.9cm,width=4.9cm]{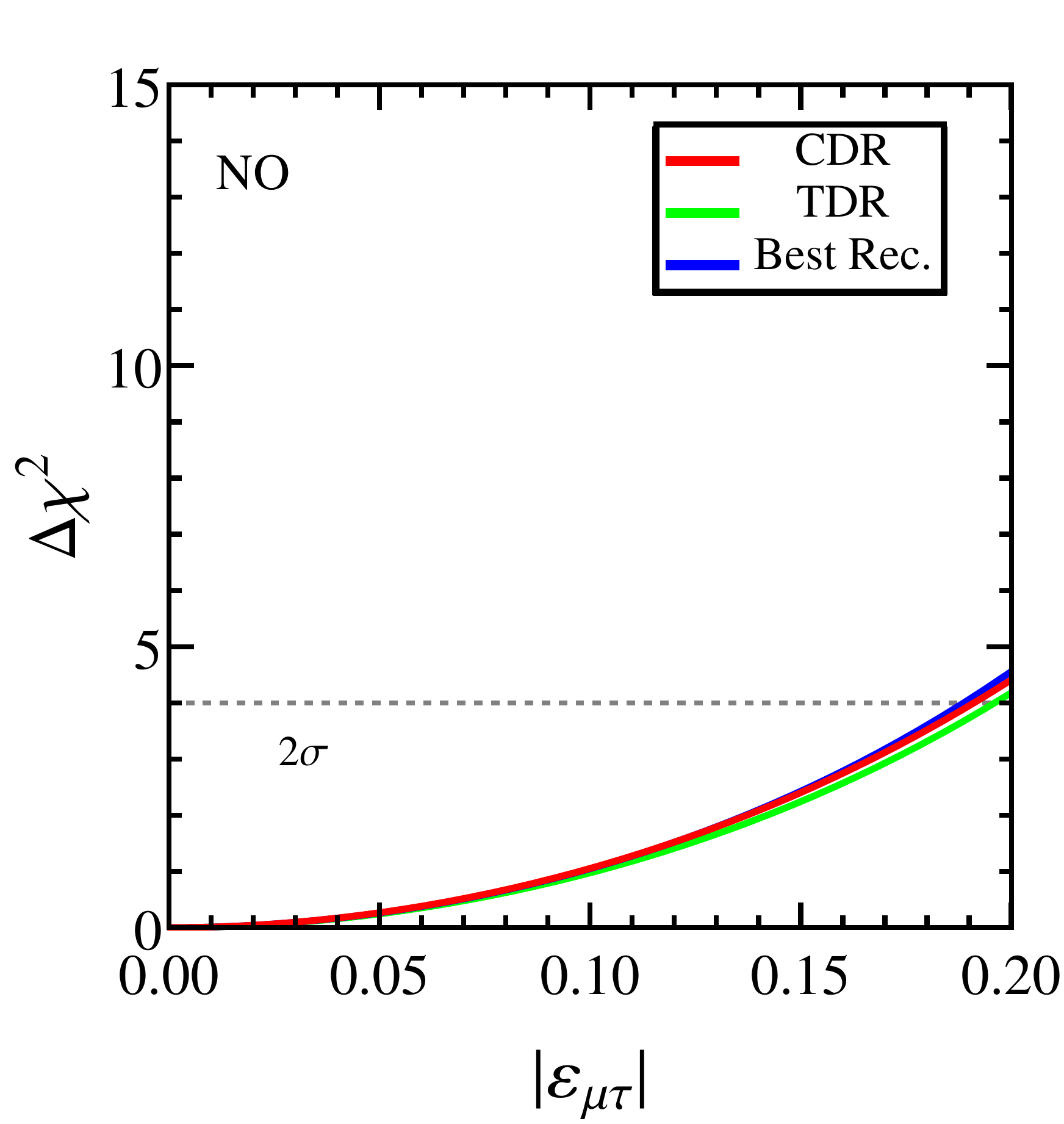}
\caption{Same as in Fig.~\ref{chisq_proj} for the non-diagonal NSI parameters $\varepsilon_{e\mu}$, $\varepsilon_{e\tau}$, and $\varepsilon_{\mu\tau}$, but now assuming them to be complex and marginalizing over their phases. }
\label{chisq_proj_nd}
\end{figure}

We show DUNE's sensitivity to complex non-diagonal NSI parameters in Fig.~\ref{chisq_proj_nd}.
While the procedure adopted here is the same as in the previous figure, we have additionally marginalized over the corresponding NSI $CP$ phases.
As discussed, the sensitivity to $|\varepsilon_{\mu\tau}|$ greatly diminishes (right panel), while a smaller but still significant effect can be seen for $|\varepsilon_{e\tau}|$ (middle panel).
$|\varepsilon_{e\mu}|$ is largely unaffected by the presence of a nonzero phase.
Even accounting for the $CP$ phases, the improvement due to a better energy resolution remains for $\varepsilon_{e\tau}$, as we can see in the middle panel of Fig.~\ref{chisq_proj_nd}.
Quantitatively, we found the upper bounds on $|\varepsilon_{e\tau}|$ at 2$\sigma$ ($\Delta\chi^2=4$) to be 0.12, 0.10 and 0.08 for the CDR, TDR and Best Reconstruction cases, respectively, revealing a possible 20\% improvement between the last two.
In Table~\ref{table2}, we quote the $2\sigma$ sensitivity on the non-diagonal, complex NSI parameters for all three scenarios.
We highlight that DUNE will significantly improve current sensitivities to $\varepsilon_{e\mu}$ and $\varepsilon_{e\tau}$ (for both real and complex NSI cases).
For the $\varepsilon_{\mu\tau}$, DUNE could compete with the IceCube constraint if the NSI is real. 
The presence of a nonzero $\phi_{\mu\tau}$ considerably worsens DUNE sensitivity, but not IceCube's~\cite{IceCube:2021abg}.


%
\begin{table*}[t!]
{%
\newcommand{\mc}[3]{\multicolumn{#1}{#2}{#3}}
\newcommand{\mr}[3]{\multirow{#1}{#2}{#3}}
\begin{center}
\begin{tabular}{|c|c|c|c|}\hline
 NSI Parameter & CDR & TDR & Best Rec. \\
\hline\hline
\mr{2}{*}{$|\varepsilon_{e\mu}|$}& \mr{2}{*}{$\leq 0.048$} &\mr{2}{*}{$\leq 0.052 $} & \mr{2}{*}{$\leq 0.047 $} \\
 &  & & \\
\hline
\mr{2}{*}{$|\varepsilon_{e\tau}|$}& \mr{2}{*}{$\leq 0.123 $} &\mr{2}{*}{$\leq 0.096 $} & \mr{2}{*}{$\leq 0.085 $} \\
 &  &&\\
\hline
\mr{2}{*}{$|\varepsilon_{\mu\tau}|$}& \mr{2}{*}{$\leq 0.191 $} &\mr{2}{*}{$\leq 0.196 $} & \mr{2}{*}{$\leq 0.189 $} \\
 &   && \\ \hline
\end{tabular}
\end{center}
}%
\caption{Same as in Table~\ref{table1}, but for the magnitude of the non-diagonal NSI parameters which are taken to be complex here (cf.~Fig.~\ref{chisq_proj_nd}).
}
\label{table2}
\end{table*}
%
%
%

\begin{figure}[t!]
\centering
\includegraphics[height=4.9cm,width=4.9cm]{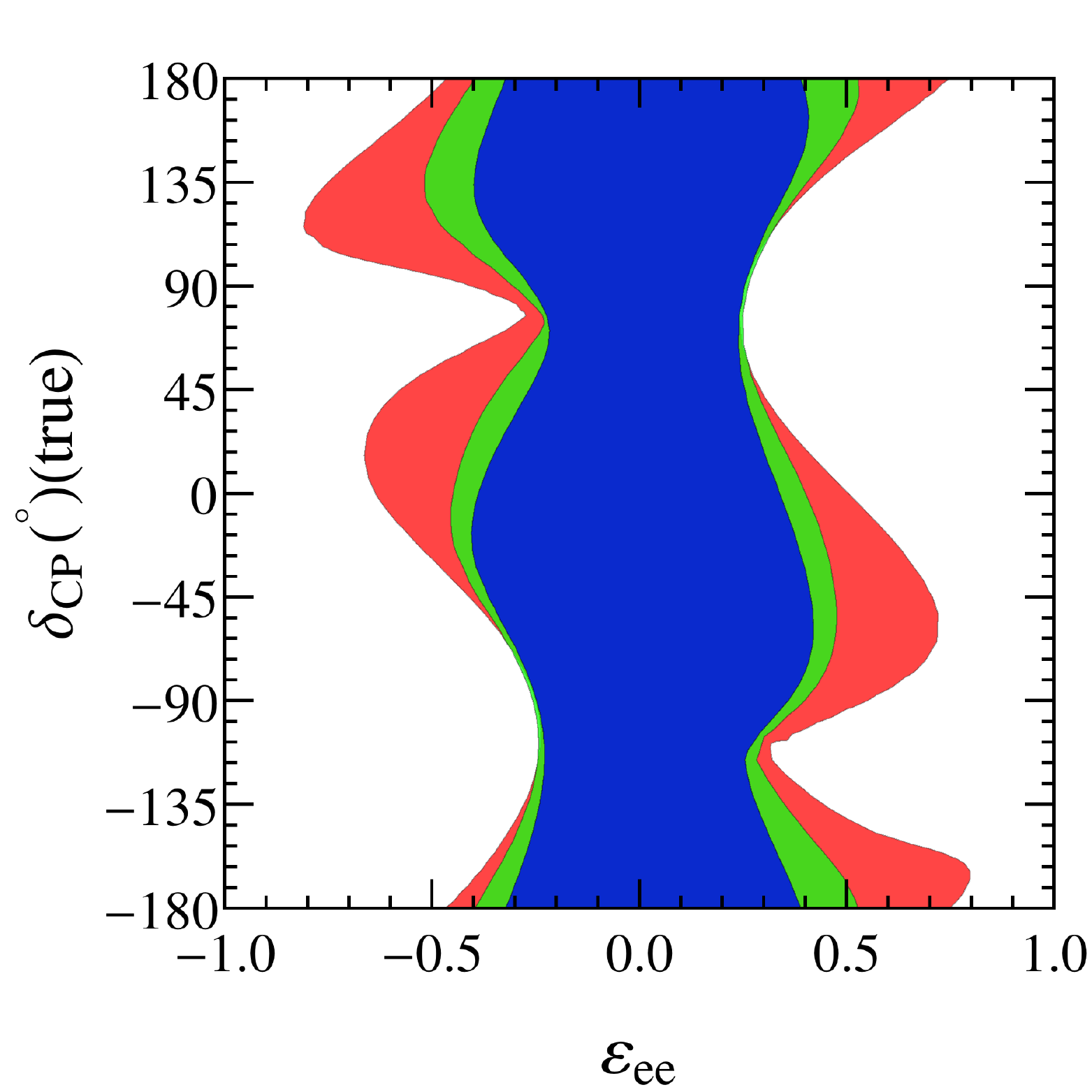}
\includegraphics[height=4.9cm,width=4.9cm]{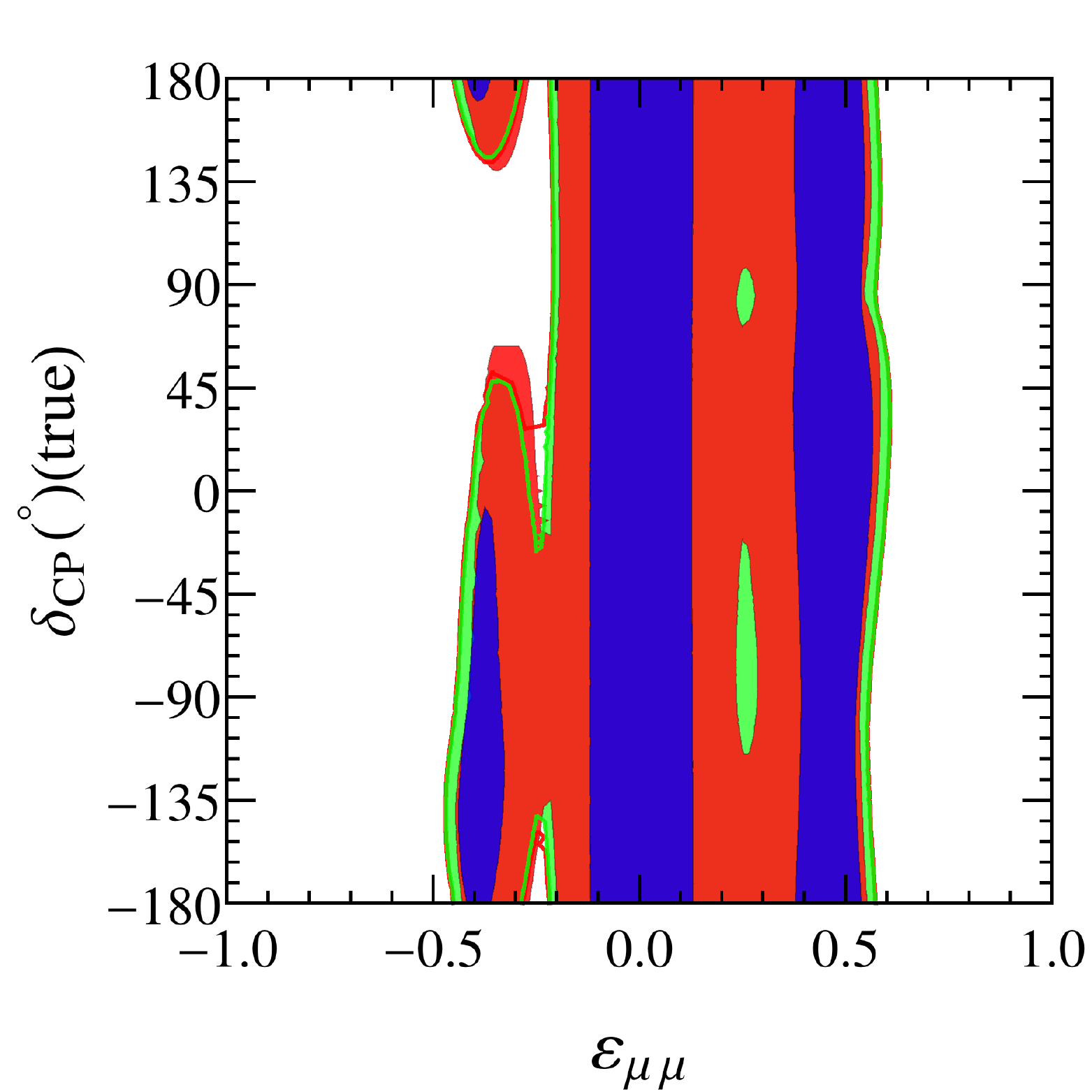}
\includegraphics[height=4.9cm,width=4.9cm]{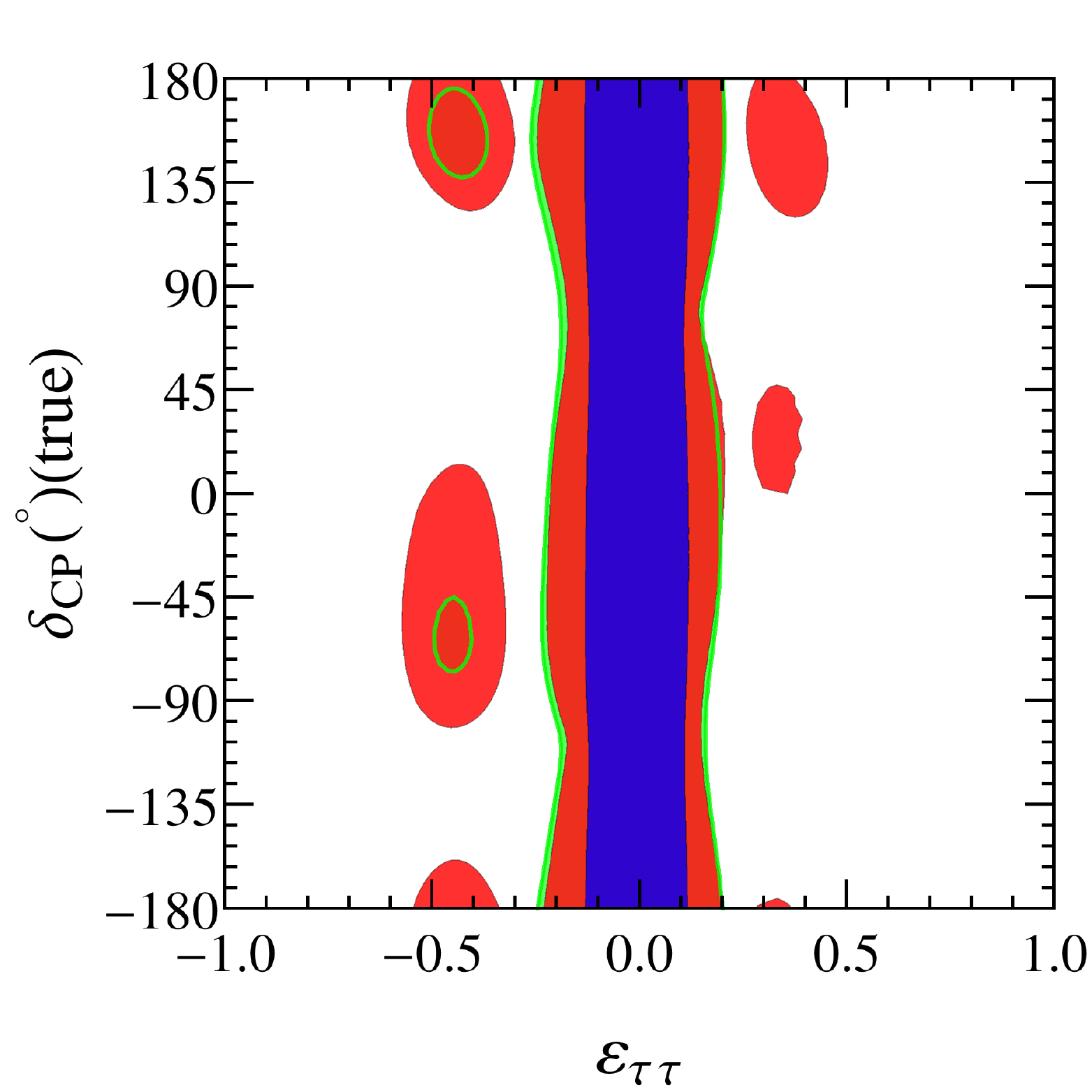}\\
\includegraphics[height=4.9cm,width=4.9cm]{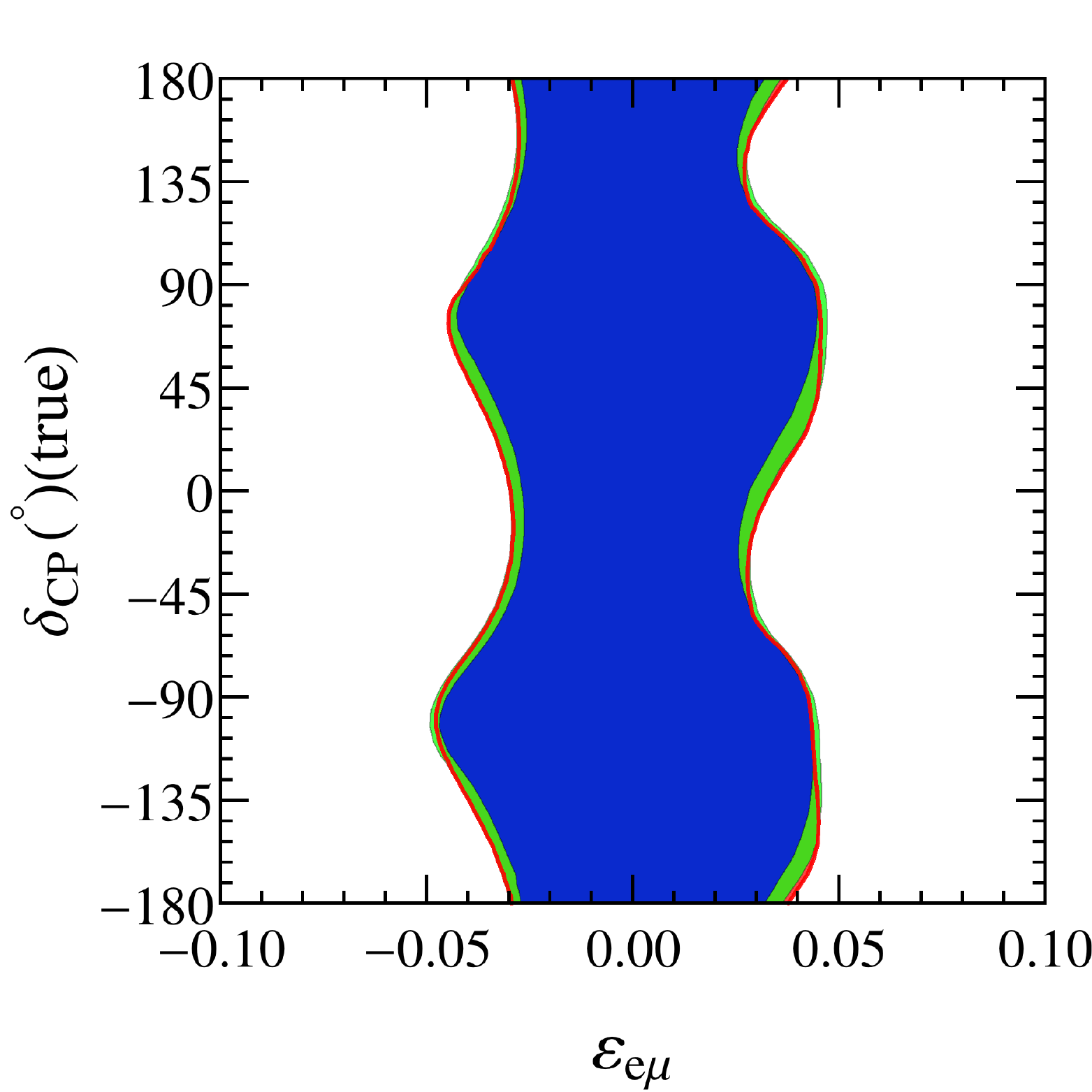}
\includegraphics[height=4.9cm,width=4.9cm]{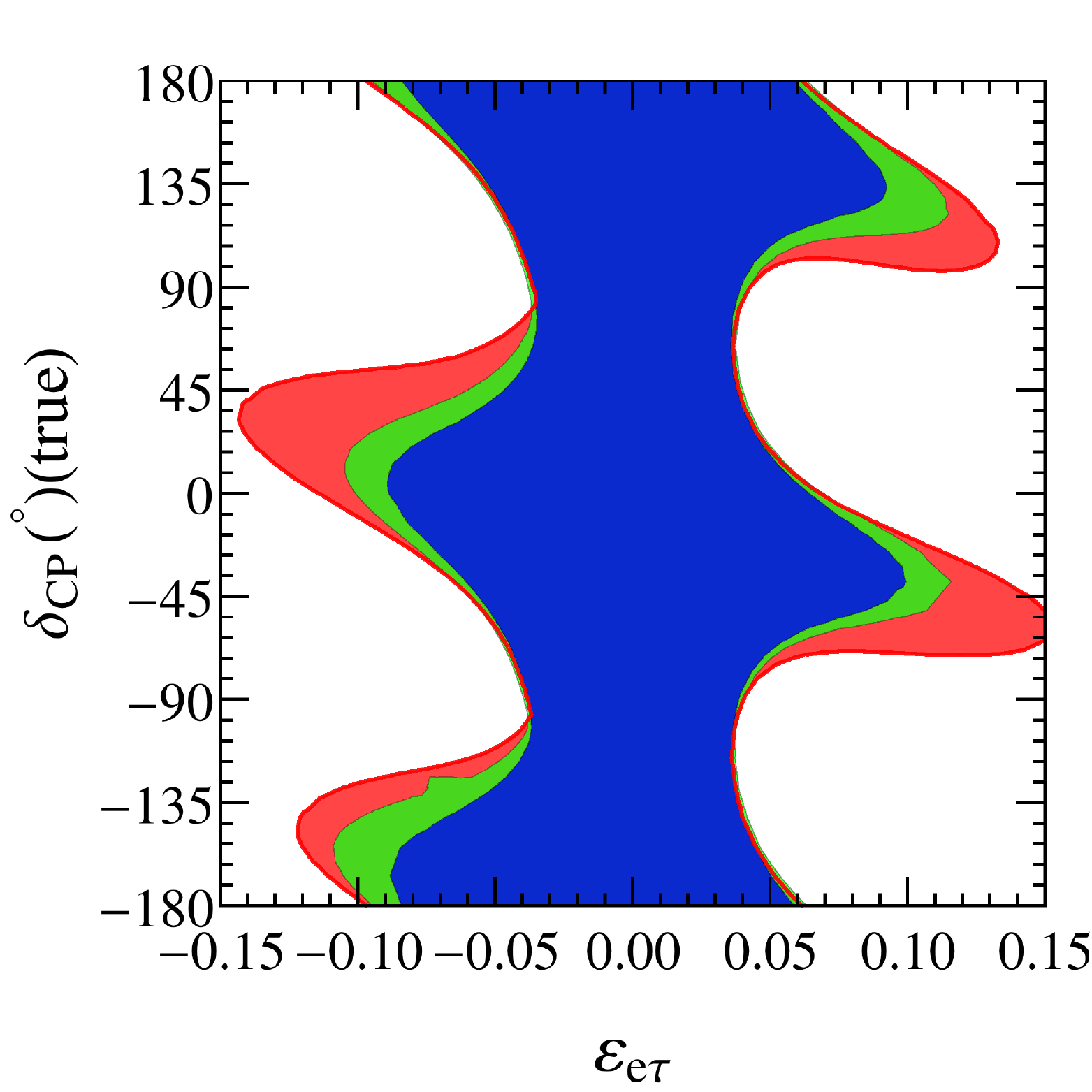}
\includegraphics[height=4.9cm,width=4.9cm]{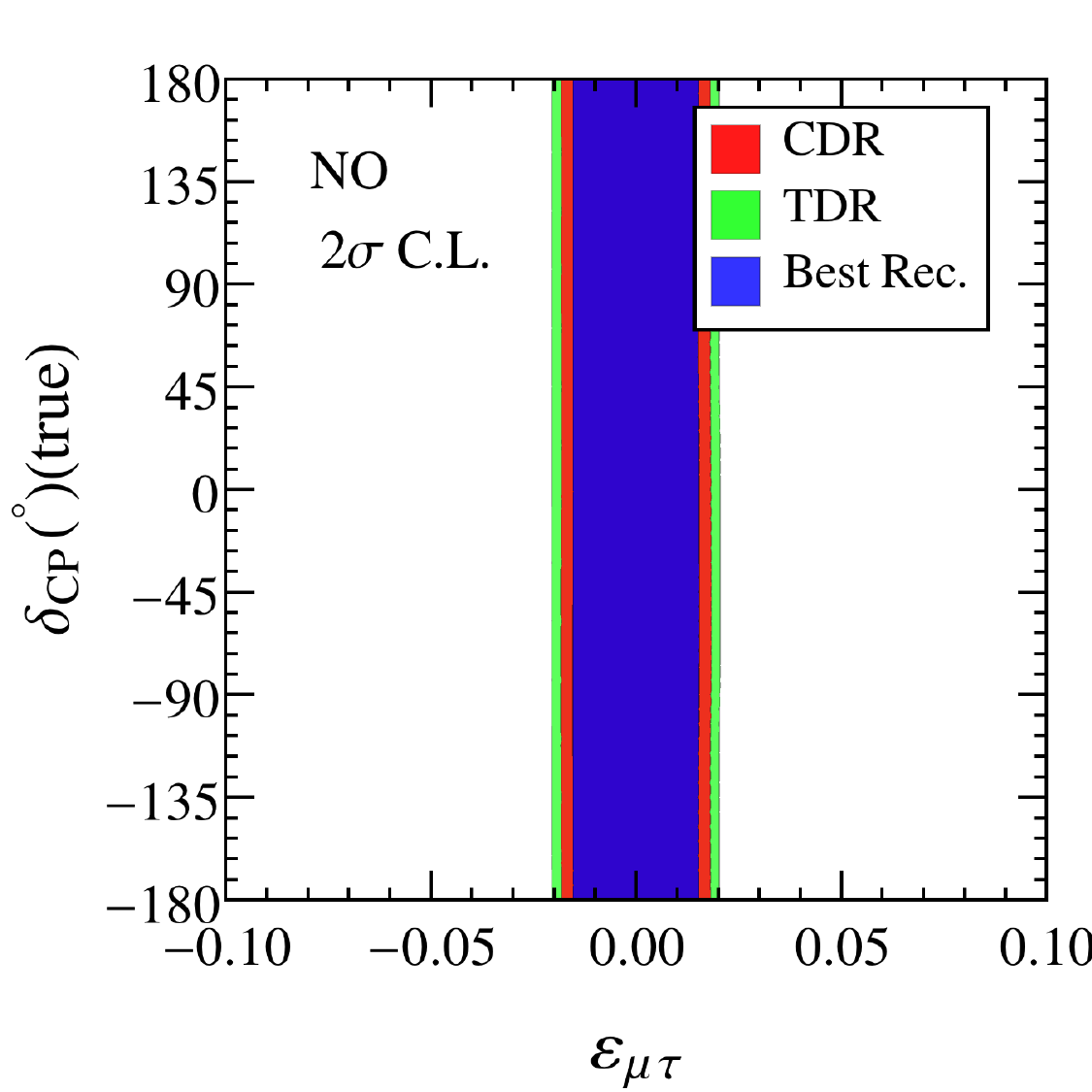}
\caption{Projected allowed regions in the plane spanned by the $\delta_{\rm CP}$\,(true) and the NSI parameters at $2\sigma$ C.L. ($\Delta\chi^2\,=\,4$). The red contours represent the DUNE sensitivity for the CDR configuration, green contours correspond to the TDR setup, and the blue contours are for the Best Reconstruction case. Normal mass ordering is assumed to be fixed in the data as well as in theory. We have considered the NSI parameters to be real. }
\label{contr_eps_delcp1}
\end{figure}

\begin{figure}[t!]
\includegraphics[height=4.9cm,width=4.9cm]{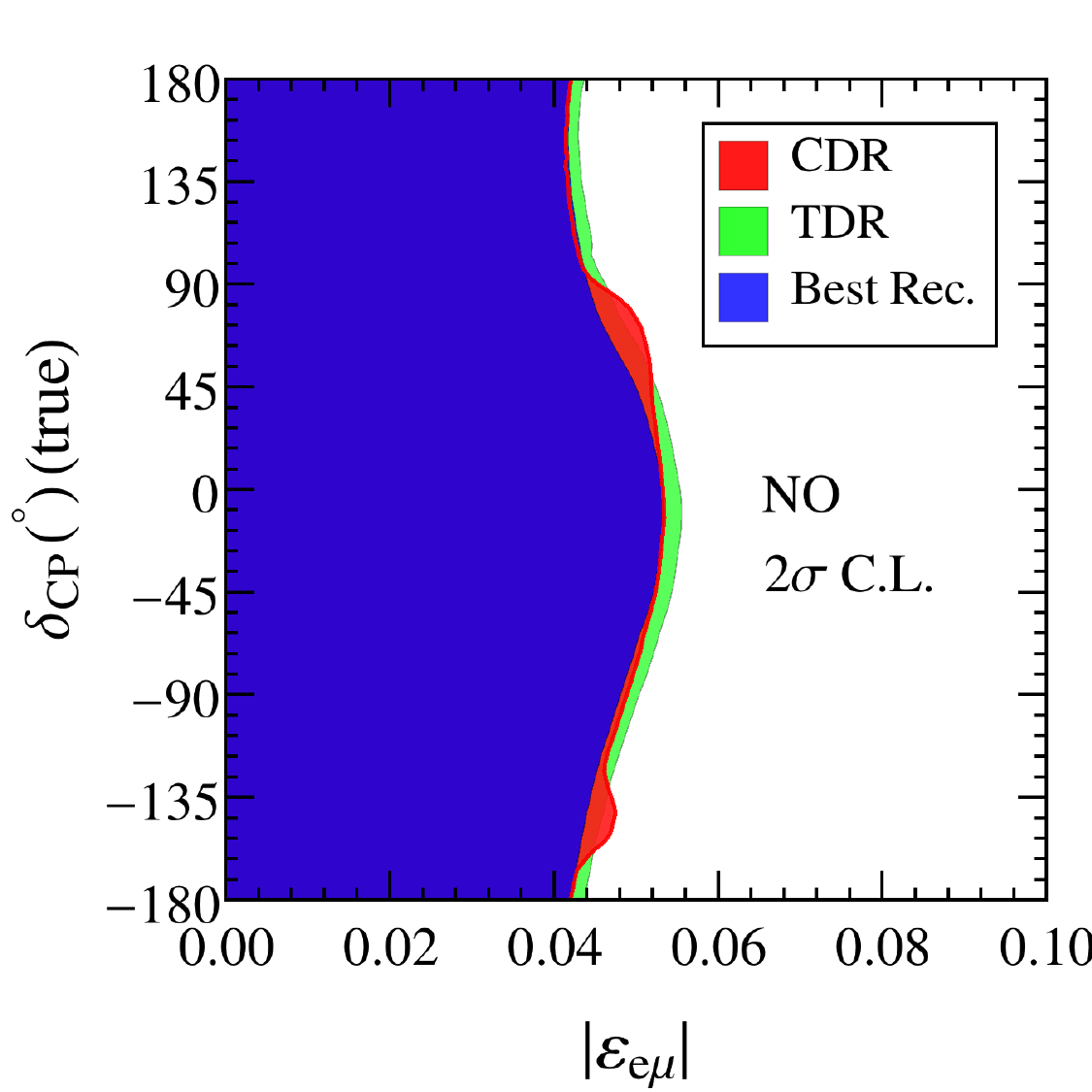}
\includegraphics[height=4.9cm,width=4.9cm]{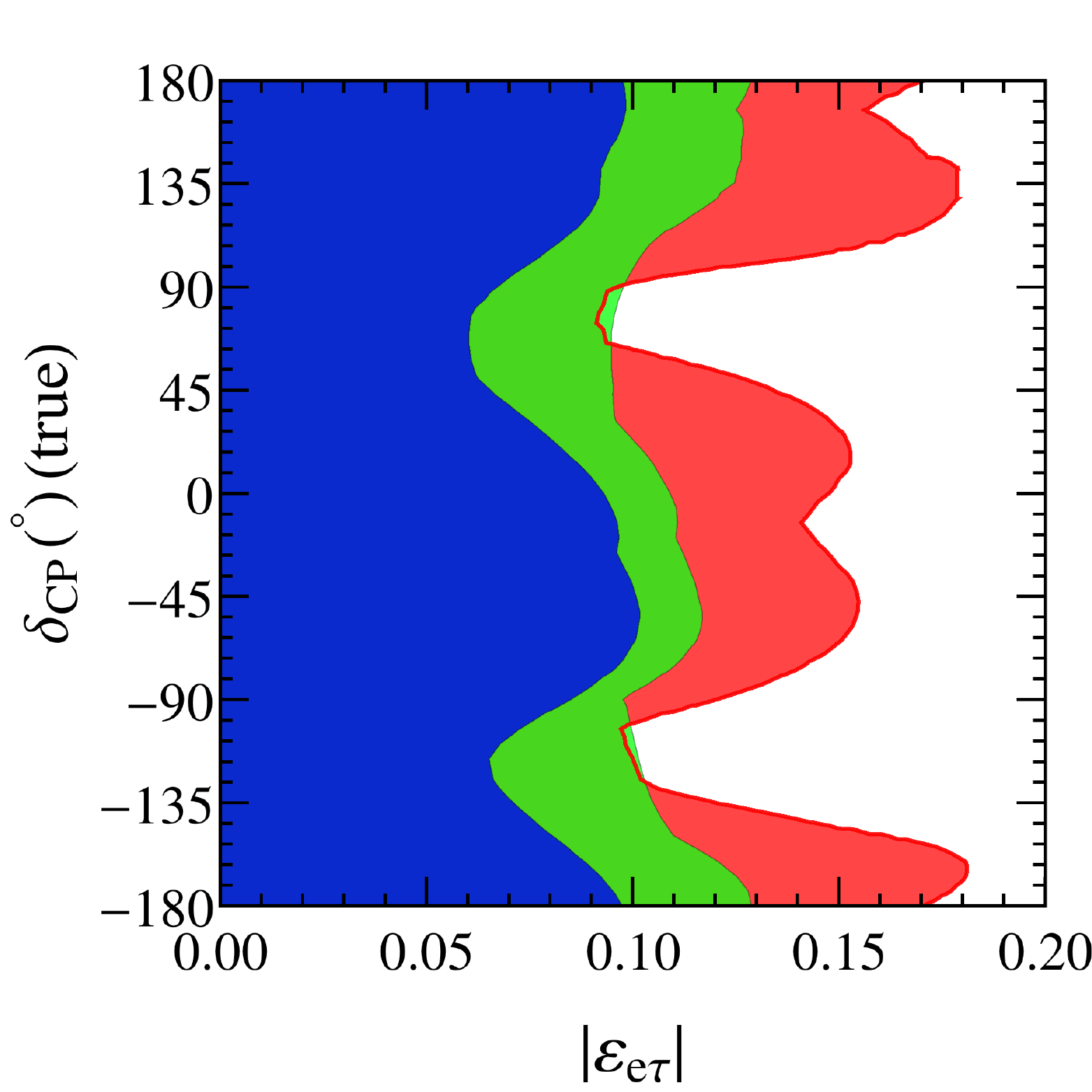}
\includegraphics[height=4.9cm,width=4.9cm]{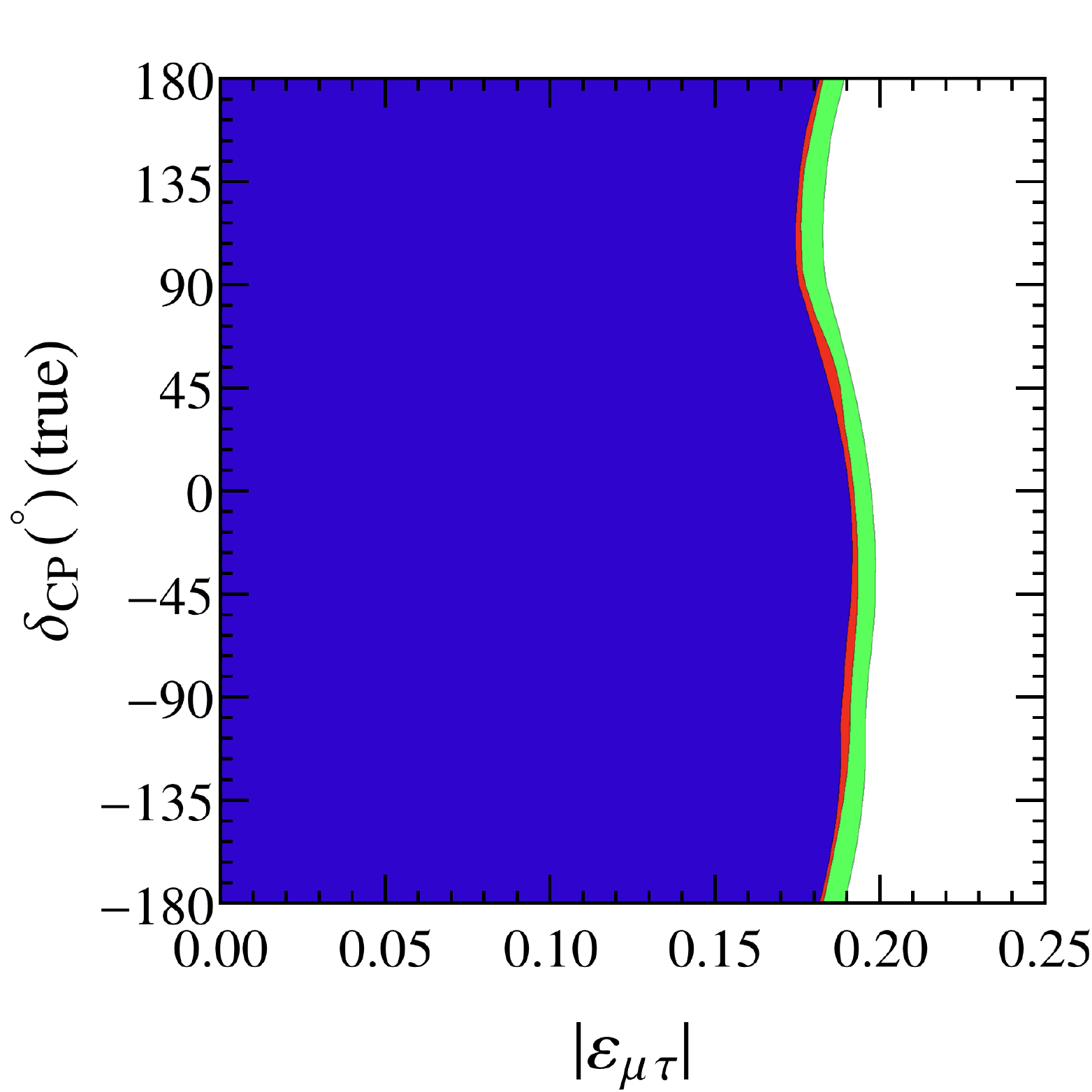}
\caption{Same as in Fig.~\ref{contr_eps_delcp1} for the non-diagonal NSI parameters, but now assuming them to be complex. }
\label{contr_eps_delcp2}
\end{figure}

To show how an improved energy resolution can affect  the sensitivity to NSIs for different  true values of the $CP$ violation phase $\delta_{\rm CP}$, we present Figs.~\ref{contr_eps_delcp1} and \ref{contr_eps_delcp2}.
The simulation details are the same as used before.
The contours in each panel is shown for $2\sigma$ C.L. with 1 d.o.f. (i.e.~$\Delta\chi^2 = 4$). The CDR, TDR and Best Reconstruction cases are represented by red, green and blue contours respectively. 
Fig.~\ref{contr_eps_delcp1} shows the allowed regions for each NSI parameter ($\varepsilon_{ee},\varepsilon_{e\mu}, \varepsilon_{e\tau}, \varepsilon_{\mu\tau}, \varepsilon_{\mu\mu},\, \textrm{and}\, \varepsilon_{\tau\tau}$), taking one NSI at a time and assuming all NSI parameters to be real.
The null hypothesis corresponds to no NSI and a given value of $\delta_{\rm CP}(\textrm{true})$ as indicated on the $y$-axis.
Therefore, this figure should be understood as the allowed region of each NSI parameter for a given input value of $\delta_{\rm CP}$. 
The variation among CDR, TDR and Best Reconstruction shows the role of the energy resolution.
If we take $\varepsilon_{ee}$ as an example, we can see that the energy resolution can significantly improve DUNE's sensitivity to this NSI, except for $\delta_{\rm CP}$(true)$\sim 80^\circ, -110^\circ$.
It is evident from this figure that the better energy resolution in DUNE plays a significant role in improving the constraints on most of the NSI parameters. 
Also note that some spurious degeneracies in the $\varepsilon_{\tau\tau}$ case are only present for the CDR and TDR resolutions, while completely disappear for the Best Reconstruction scenario.

In the previous discussion we focused on real NSI parameters. 
Now, we present in Fig.~\ref{contr_eps_delcp2} the allowed region for the absolute value of non-diagonal NSI parameters, marginalizing over their corresponding $CP$ phase, as a function of the true value of $\delta_{\rm CP}$.
All analysis details are the same as before.
As expected, the allowed regions are larger when one consider complex NSI parameters, particularly for $\varepsilon_{\mu\tau}$ due to the $\cos\phi_{\mu\tau}$ factor in the leading-order contribution of this NSI to the $\nu_\mu$ survival probability, see Eq.~\eqref{eq:Pmm}.
We also observe that the energy resolution plays an important role in the sensitivity to $\varepsilon_{e\tau}$ even in the presence of the NSI $CP$ phase.

\begin{figure}[t!]
\centering
\includegraphics[width=0.328\textwidth]{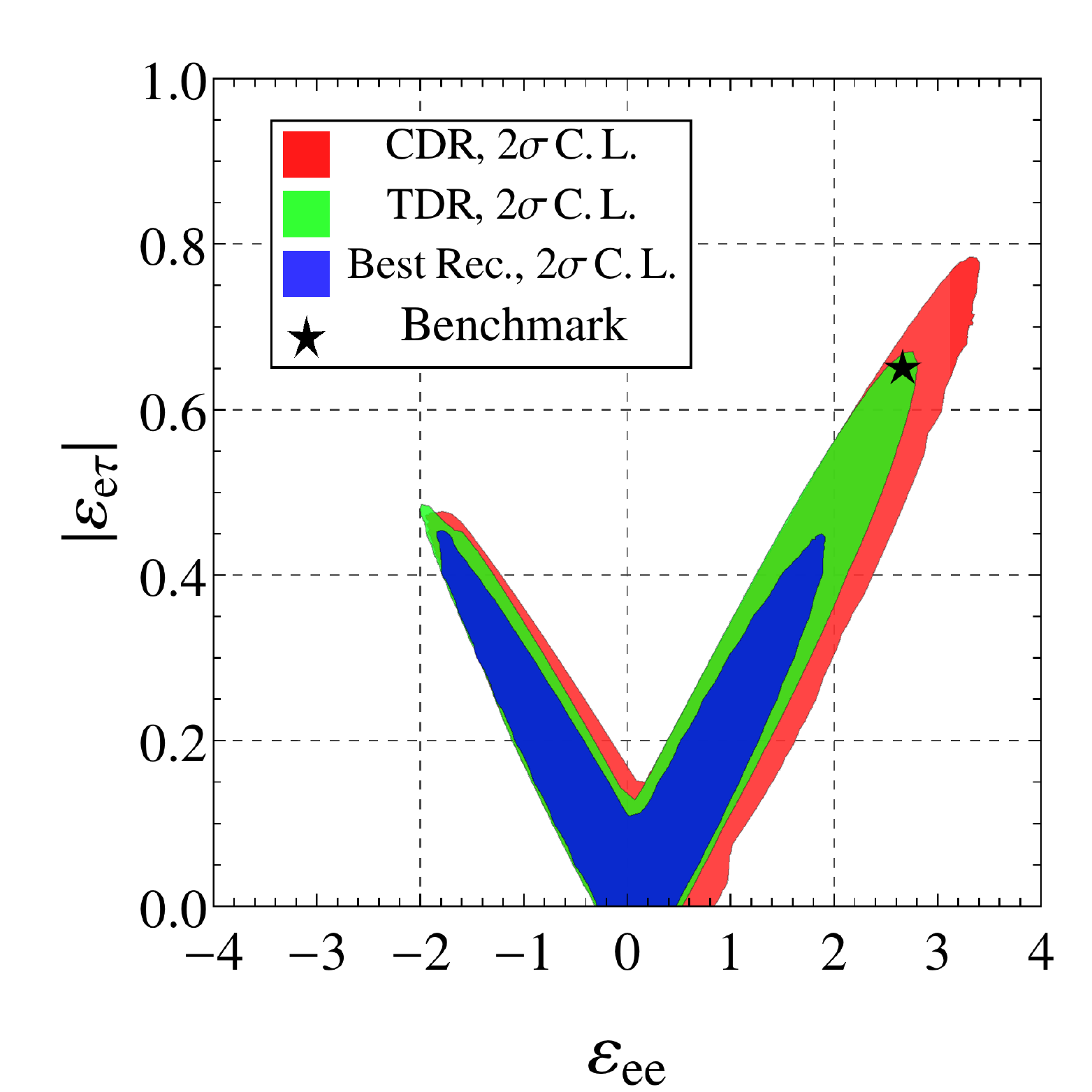}
\includegraphics[width=0.328\textwidth]{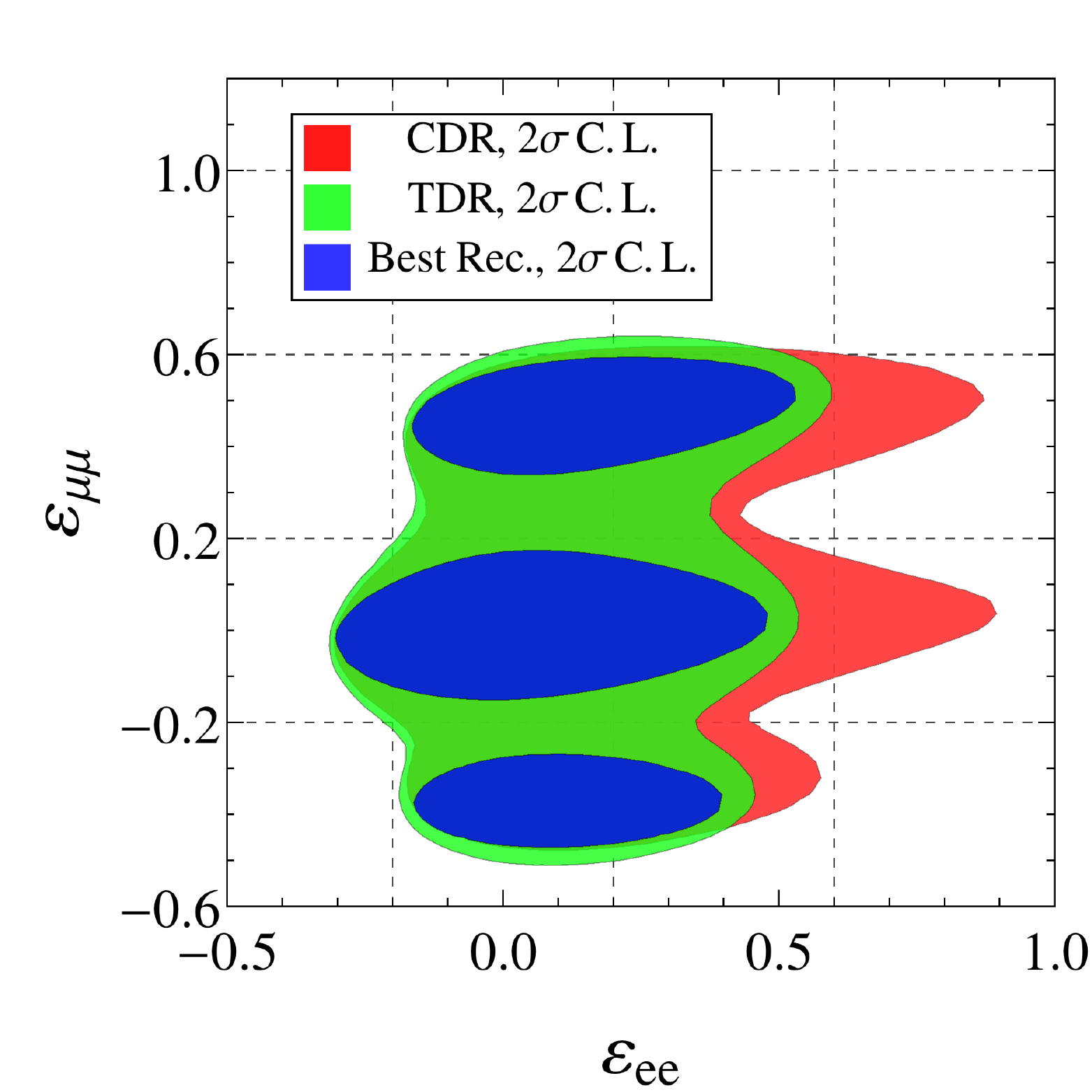}
\includegraphics[width=0.328\textwidth]{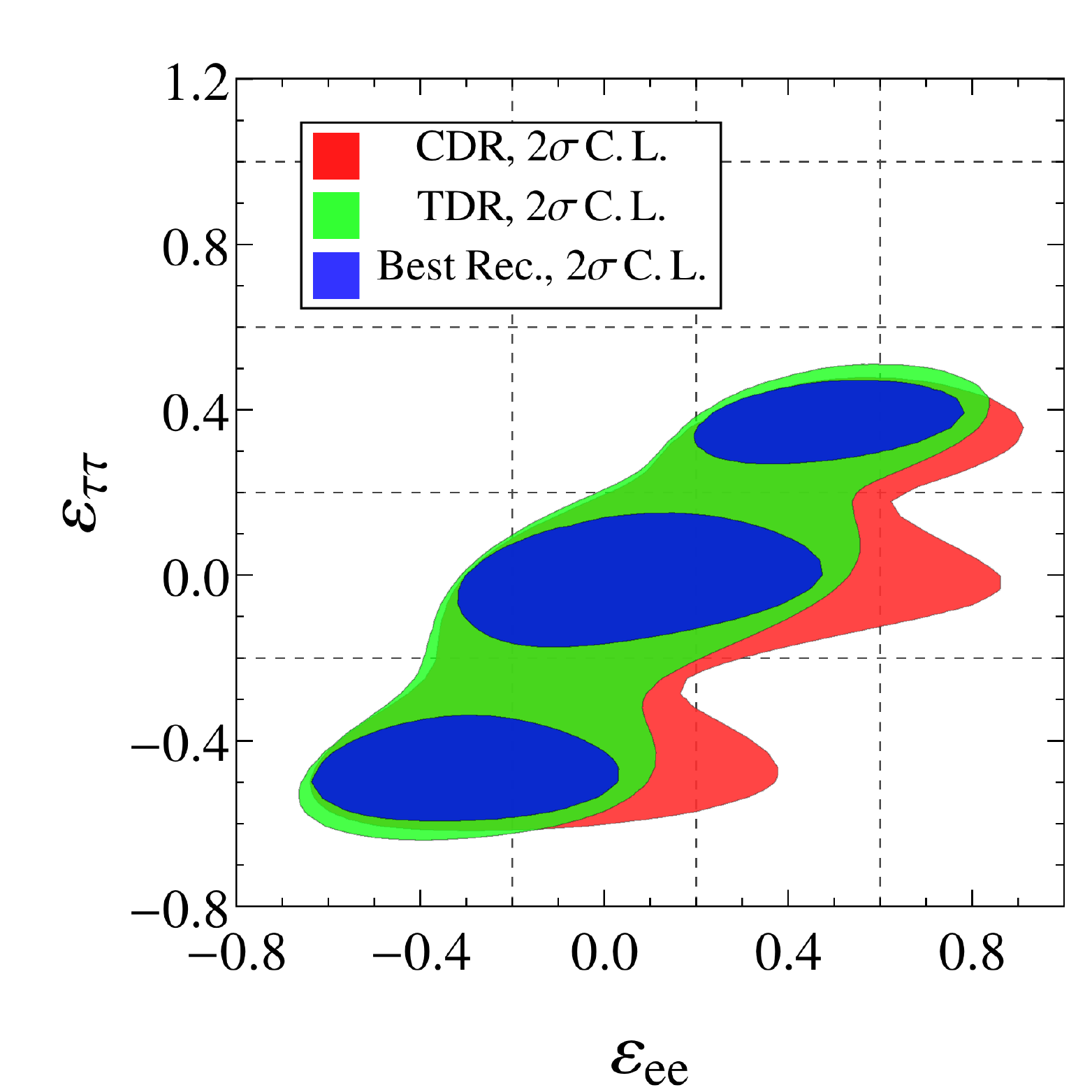}\\
\includegraphics[width=0.328\textwidth]{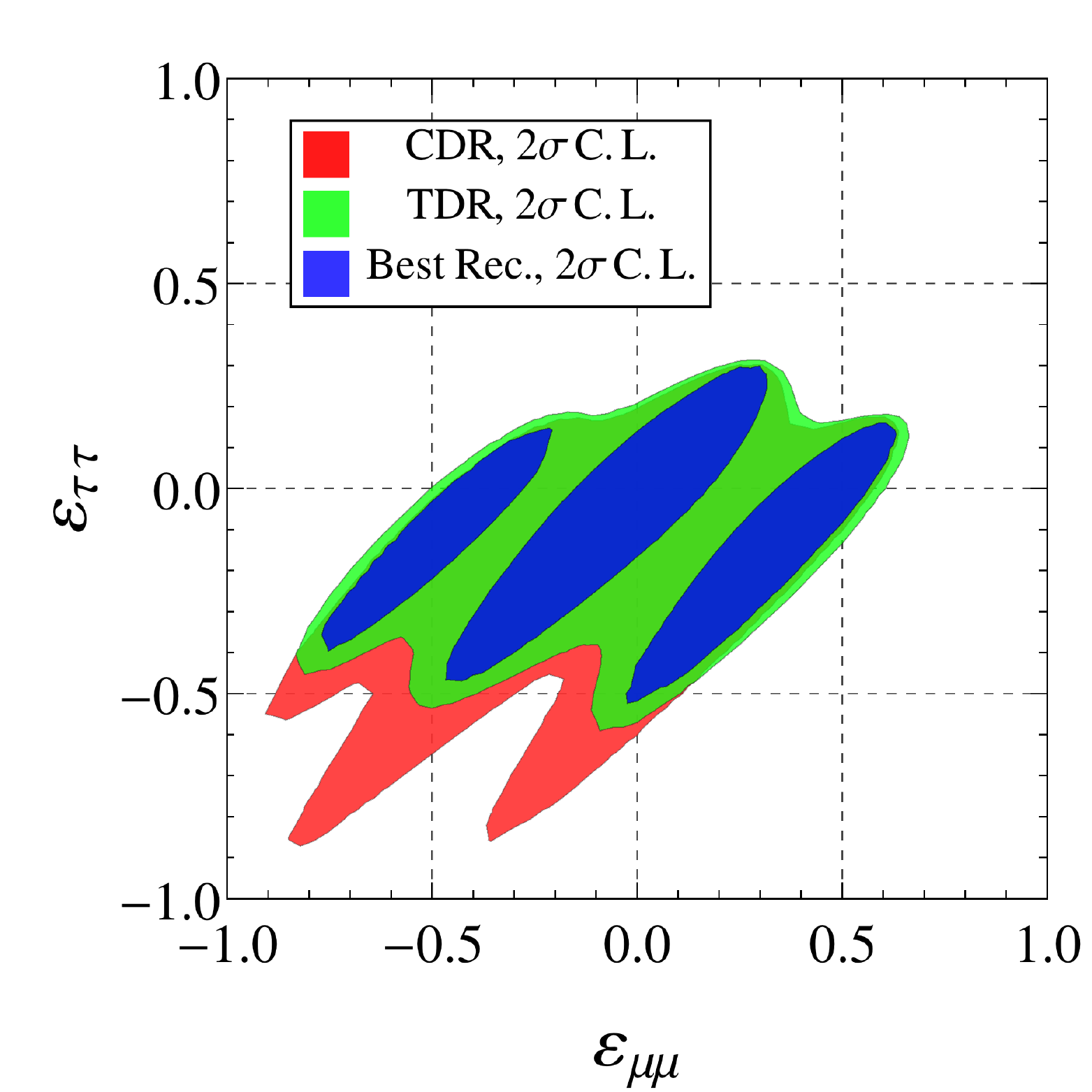}
\includegraphics[width=0.328\textwidth]{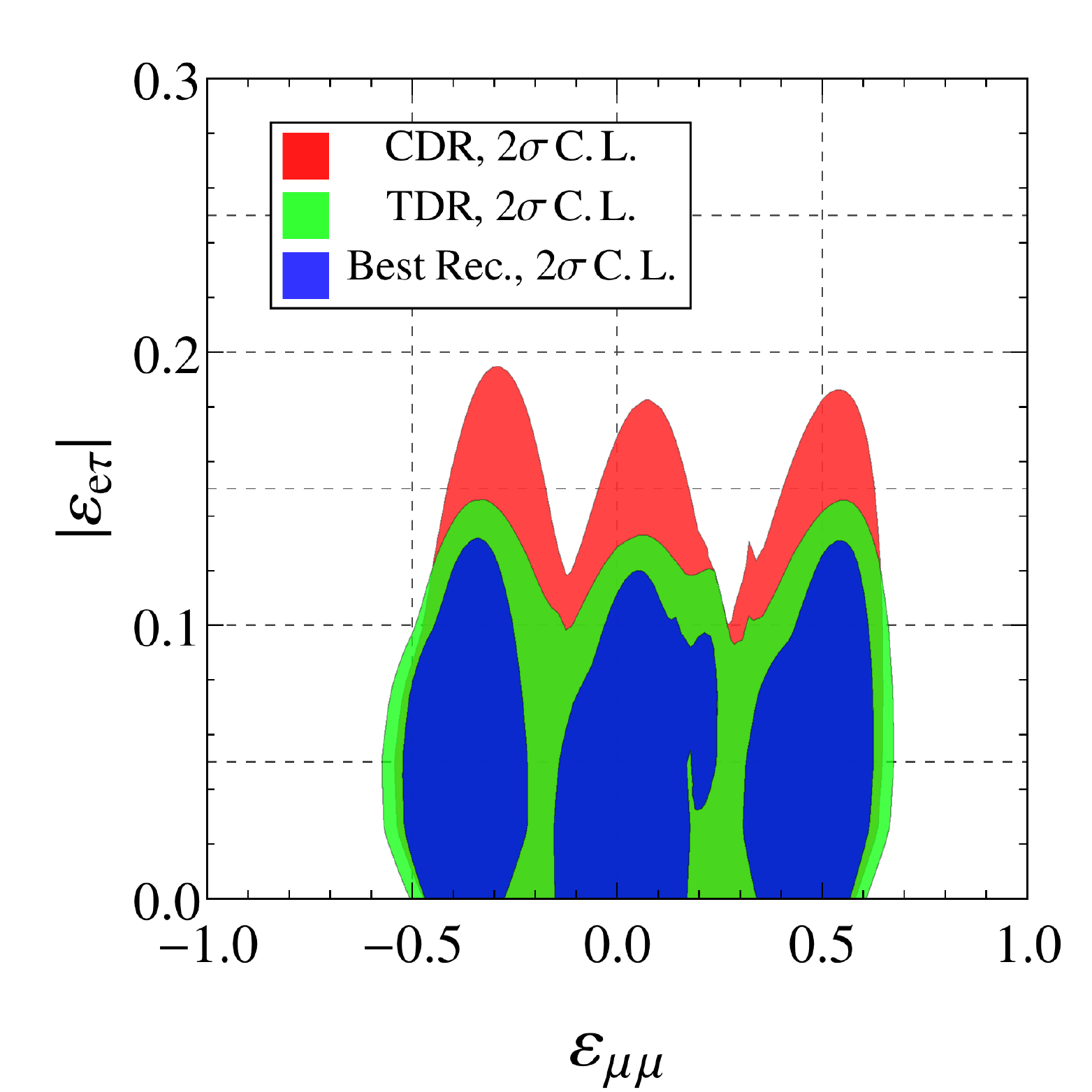}
\includegraphics[width=0.328\textwidth]{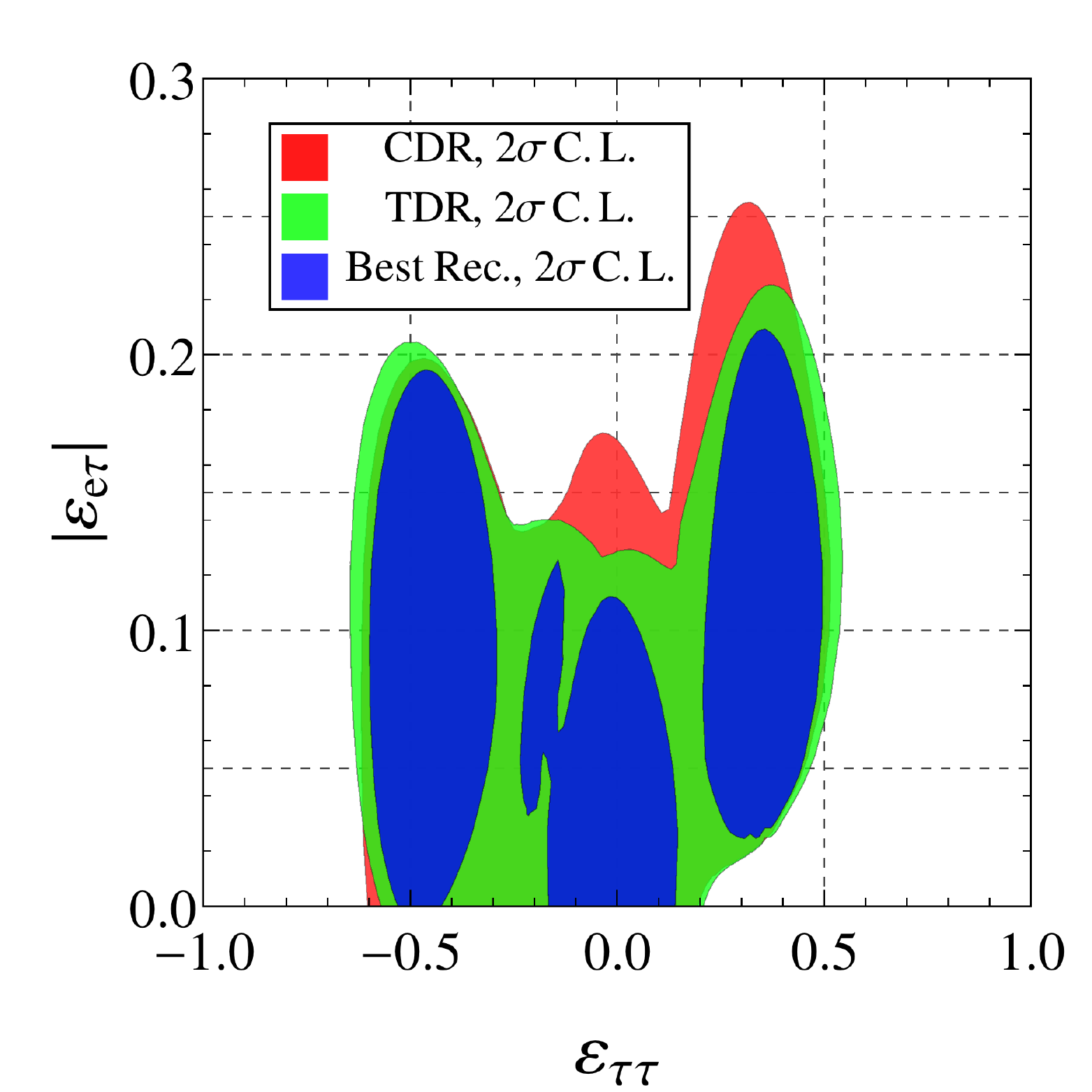}
\caption{$2\sigma$ C.L. (2 d.o.f.) sensitivity regions spanned by the different sets of NSI parameters for three different cases CDR, TDR, and Best Reconstruction respectively. We have assumed $\delta_{\rm CP}(\rm true) = -90^{\circ}$ and normal ordering is assumed in both true and test hypotheses of the analysis. The benchmark point marked in the upper left panel will be used later in Sec.~\ref{sec:discussion}.}
\label{fig:2Dproj1}
\end{figure} 
Now that we have established the important role of the energy resolution in constraining certain NSI parameters, it is important to assess how robust these conclusions are if we do not assume one NSI parameter at a time.
In general, UV complete models will predict a combination of NSI parameters, with possible correlations stemming from the new particles and interactions that  generate the NSI at a higher scale~\cite{Farzan:2015doa, Farzan:2015hkd, Farzan:2016wym, Forero:2016ghr, Babu:2017olk, Dey:2018yht, Babu:2019mfe}.
Although a comprehensive study of specific models is beyond the scope of this paper, we can redo our analysis taking into account two NSI parameters at a time, instead of individual $\varepsilon_{\alpha\beta}$. Note that further redoing the analysis with more than two NSI parameters will not be very illuminating for our purposes.

In Fig.~\ref{fig:2Dproj1} we show the allowed regions at  $2\sigma$ C.L. with 2 d.o.f. ($\Delta \chi^2 = 6.18$) for different pairs of NSI parameters.
We consider $\varepsilon_{ee}$, $\varepsilon_{\mu\mu}$, $\varepsilon_{\tau\tau}$, and $\varepsilon_{e\tau}$, as those are the parameters that profit the most from an improved energy resolution (cf.~Figs.~\ref{chisq_proj} and \ref{chisq_proj_nd}).
As before, we indicate the CDR, TDR and Best Reconstruction configurations by the red, green and blue shaded regions, respectively. 
In order to draw the two-dimensional contours we have marginalized away the two mixing angles $\theta_{13}$ and $\theta_{23}$, the mass splitting $\Delta m^2_{31}$, the $CP$ phase $\delta_{\rm CP}$, and the matter density $\rho$, within the uncertainties discussed in Sec.~\ref{sec:simulation}. 
Moreover the nonstandard $CP$ phases for non-diagonal NSIs have also been marginalized away. 
We have fixed $\delta_{\rm CP}$(true) at $-90^{\circ}$ as well as normal ordering for the mass spectrum. 
We have considered a few combinations of two NSI parameters at a time keeping others fixed to zero. 
As expected (see e.g. Refs.~\cite{Coloma:2016gei, Liao:2016hsa}), we observe significant degeneracies between NSI parameters, particularly for the pairs containing $(\varepsilon_{ee},\,|\varepsilon_{e\tau}|)$.
We also see here a substantial impact of the improved energy resolution in several of these NSI combinations. 
In the upper left panel of Fig.~\ref{fig:2Dproj1} we can see that the allowed region in the Best Reconstruction scenario in the $(\varepsilon_{ee},\,|\varepsilon_{e\tau}|)$ plane for positive $\varepsilon_{ee}$ is reduced to about half of the allowed region in the CDR case.
Moreover, the allowed regions in both $\varepsilon_{\mu\mu}$ and $\epsilon_{\tau\tau}$ versus other parameters are connected for the CDR and TDR cases, whereas three disconnected islands appear for the Best Reconstruction case.
%
%
%
%
%

\begin{figure}[t!]
\centering
\includegraphics[height=7.3cm,width=7.3cm]{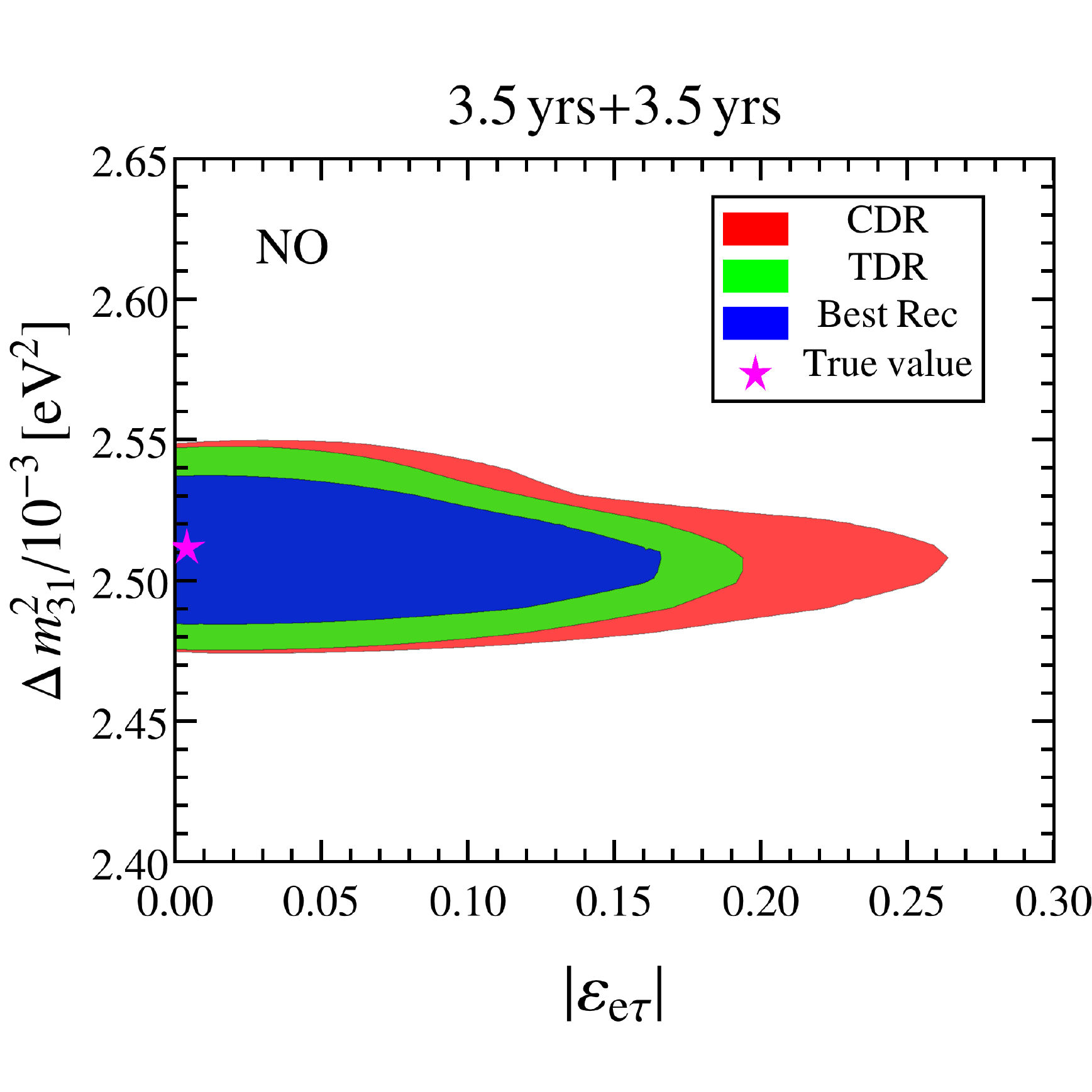}
\includegraphics[height=7.3cm,width=7.3cm]{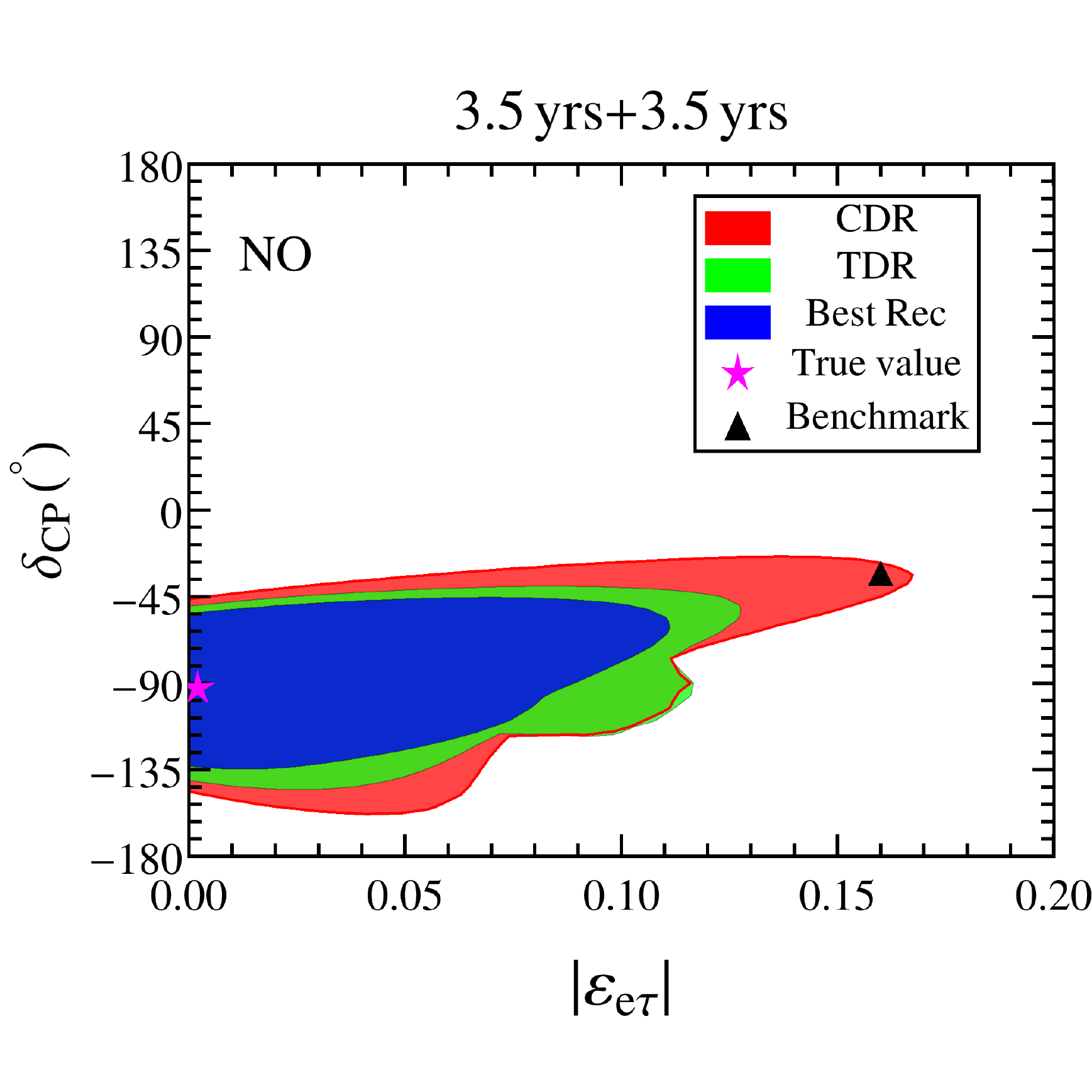}
\caption{Allowed regions in the $\varepsilon_{e\tau}$ versus $\Delta m^2_{31}$ (left) and $\varepsilon_{e\tau}$ versus $\delta_{\rm CP}$ (right) planes for the DUNE CDR, TDR and Best Reconstruction cases (red, green, and blue contours, respectively) at $2\sigma$ C.L. ($\Delta\chi^2 = 6.18$). The assumed true value is indicated as a magenta star while the black triangle on the right panel is a NSI benchmark, see text.}
\label{fig:delcp_m31_epset_new}
\end{figure} 

\section{Discussion}
\label{sec:discussion}
In this section we discuss the results obtained previously and the role of the energy resolution in a qualitative manner.
First, we present Fig.~\ref{fig:delcp_m31_epset_new}, in which we show an illustrative example of DUNE's sensitivity to NSI for CDR, TDR and Best Reconstruction (red, green and blue, resp.) in the planes of $(\varepsilon_{e\tau}, \Delta m^2_{31})$ (left panel) and $(\varepsilon_{e\tau},\delta_{\rm CP})$ (right panel)  at $2\sigma$ C.L. for 2 d.o.f. ($\Delta\chi^2\,=\,6.18$).
The magenta star in each panel represents the choice of true parameters where we have assumed no NSI.
Note that we have marginalized away the NSI $CP$ phase $\phi_{e\tau}$ in both panels.

The improvement on the measurement of $\Delta m^2_{31}$ due to a better energy resolution is obvious: a more precise determination of the minimum of oscillation allows for a better measurement of the oscillation frequency and therefore of the mass splitting.
In fact, the better energy resolution makes the first oscillation minimum in the disappearance channel more prominent and helps resolving the second oscillation minimum, around 0.8~GeV.
Nevertheless, the right panel is more interesting for us, as we see a degeneracy between $\delta_{\rm CP}$ and $\varepsilon_{e\tau}$. 
To understand the role of the energy resolution, let us take the NSI benchmark point denoted by the black triangle in the right panel, which is allowed by the CDR but excluded by the TDR or Best Reconstruction simulations.
The relevant parameters for this benchmark point are $\delta_{\rm CP}\,=\, -35^{\circ}$, $|\varepsilon_{e\tau}|\, = \, 0.16$, and $\phi_{e\tau}\, = \,10^{\circ}$ for the CDR, and $\phi_{e\tau}\, = \,15^{\circ}$ for the TDR and Best Reconstruction cases.
The other standard oscillation parameters were chosen to be the best fit values in the analysis.

\begin{figure}[t!]
\hspace{0.3cm}
\includegraphics[width=1\textwidth]{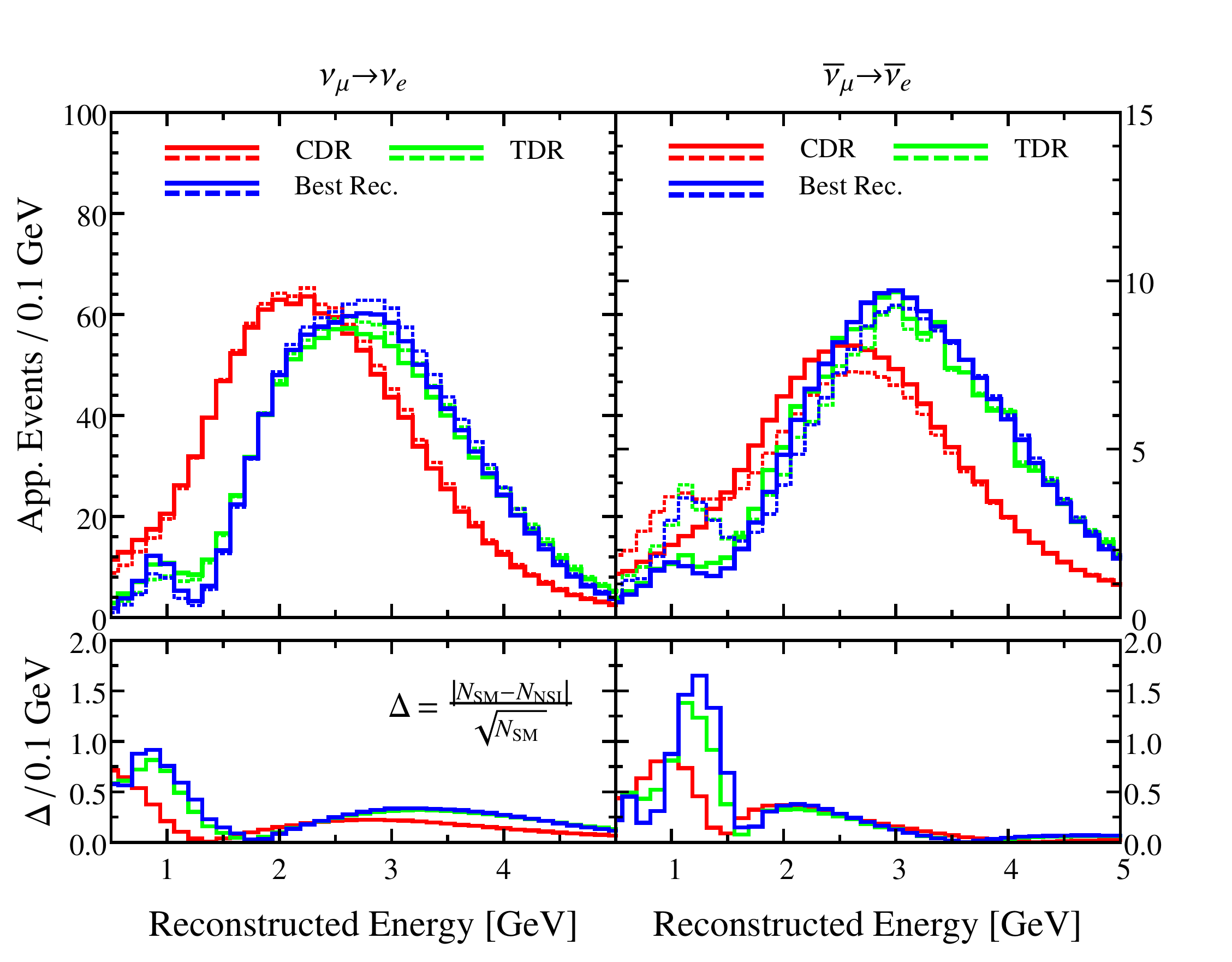}
\caption{Upper left (right) panel presents the $\nu_{\mu} \rightarrow \nu_e$ ($\bar{\nu}_{\mu} \rightarrow \bar{\nu}_e$) appearance event spectra as a function of reconstructed energy. Red, green and blue lines correspond to the CDR, TDR, and Best Reconstruction configurations, respectively. Solid lines represent the standard three-neutrino spectra with $\delta_{\rm CP}\,=\, -90^{\circ}$ and dashed lines represent the spectra with NSI in the $e-\tau$ sector. In case of NSI, we have taken the best fit parameters corresponding to the black triangle in Fig.~\ref{fig:delcp_m31_epset_new}, i.e., $\delta_{\rm CP}\,=\, -35^{\circ}$, $|\varepsilon_{e\tau}|\, = \, 0.16$, and $\phi_{e\tau}\, = \,10^{\circ}$ for the CDR, and $\phi_{e\tau}\, = \,15^{\circ}$ for the TDR and Best Reconstruction cases. The best fit values of $\theta_{13}$, $\theta_{23}$, and $\Delta m_{31}^2$ have also been chosen from the simulation. Lower left (right) panel represents the absolute difference in the upper left (right) appearance events between the standard and the NSI scenarios divided by the statistical uncertainty in each bin. The color coding is the same as in the upper panels.}
\label{fig:spec_etau}
\end{figure} 

In Fig.~\ref{fig:spec_etau} we show the appearance spectra for the neutrino mode (upper left panel) and antineutrino mode (upper right panel) for the CDR, TDR and Best Reconstruction (red, green and blue, respectively), for the assumed true parameters (magenta star in Fig.~\ref{fig:delcp_m31_epset_new} right panel, solid lines in Fig.~\ref{fig:spec_etau}) and NSI benchmark point (black triangle in Fig.~\ref{fig:delcp_m31_epset_new} right panel, dashed lines in Fig.~\ref{fig:spec_etau}). First we can understand why it is hard to disentangle this NSI effect: the spectra themselves look very similar between the standard model  assumed true values and the NSI benchmark. This is particularly true for the first maximum. 
The second maximum shows a larger deviation, but with lower statistics. To better see the differences between the standard and NSI cases, we  show in the lower panels the absolute difference  on the number of events divided by the statistical uncertainty in each bin, i.e. $\Delta =|N_{\rm SM}-N_{\rm NSI}|/\sqrt{N_{\rm SM}}$. 
This statistical figure of merit should be viewed only as a crude approximation for the statistical relevance of each bin, but it serves to demonstrate the relevance of the first and second oscillation maxima in a qualitative level.
As we can see, a better energy resolution not only leads to a slight increase in the statistical power around the first oscillation maximum, but also on the second maximum.
This draws us to the conclusion that the improved energy resolution enhances the sensitivity not only due to a better reconstruction of the first maximum: the role of the second maximum is indeed very much significant. 
These considerations apply to both neutrino and antineutrino modes.

\begin{figure}[t!]
\hspace{0.3cm}
\includegraphics[width=1\textwidth]{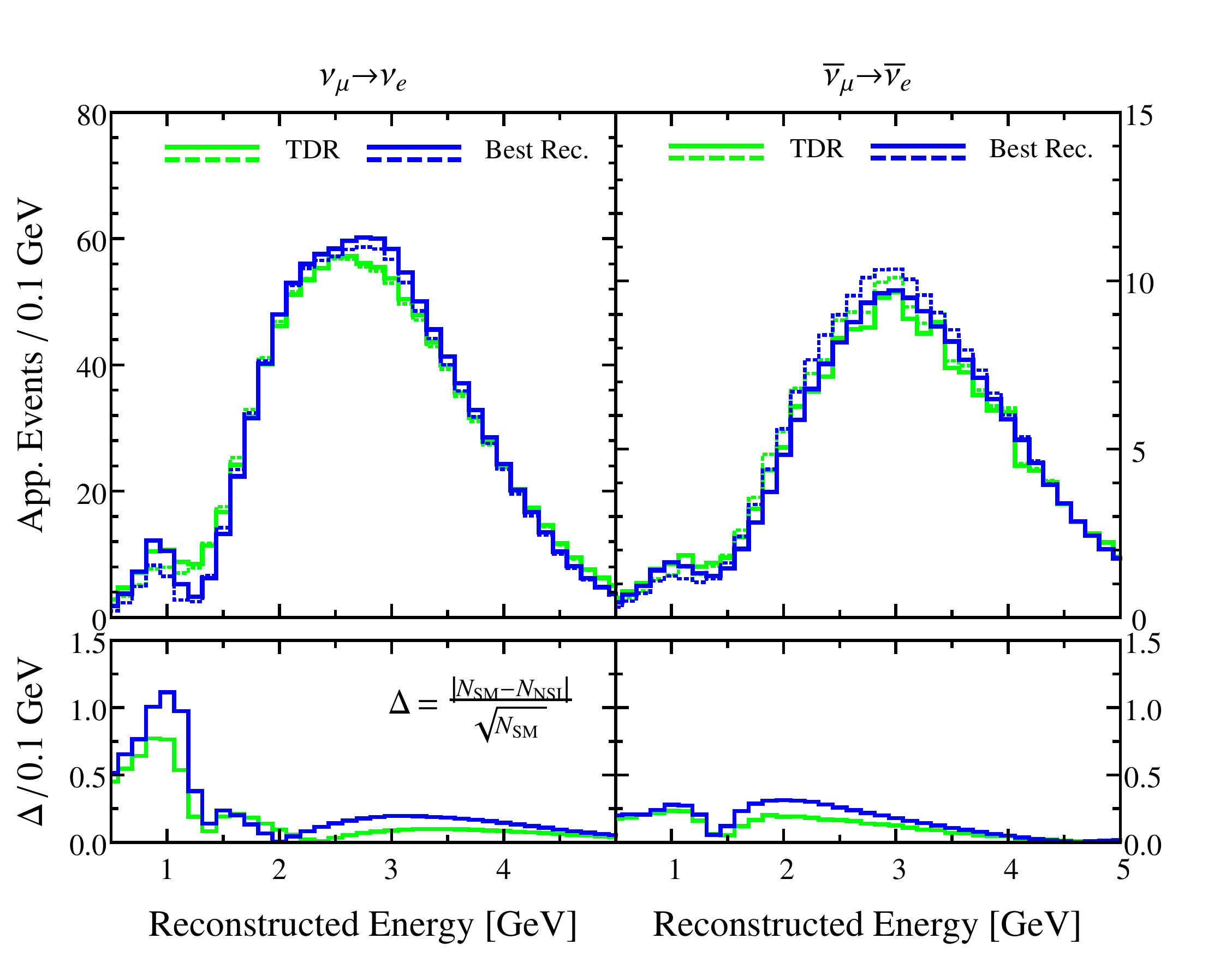}
\caption{Upper left (right) panel presents the $\nu_{\mu} \rightarrow \nu_e$ ($\bar{\nu}_{\mu} \rightarrow \bar{\nu}_e$) appearance event spectra as a function of reconstructed energy. Green and blue lines correspond to the TDR and Best Reconstruction configurations respectively. Solid lines represent the standard three-neutrino spectra with $\delta_{\rm CP}\,=\, -90^{\circ}$ and dashed lines represent the spectra with NSIs in the $e-e$ and $e-\tau$ sectors simultaneously. In case of NSI, we have taken the best fit parameters corresponding to the black star in the upper left panel of Fig.~\ref{fig:2Dproj1}, i.e., $\delta_{\rm CP}\,=\, -22^{\circ}$, $\varepsilon_{ee}\, = \, 2.70$, $|\varepsilon_{e\tau}|\, = \, 0.65$, and $\phi_{e\tau}\, = \,-177^{\circ}$ for the TDR, and $\delta_{\rm CP}\,=\, -23^{\circ}$, $\varepsilon_{ee}\, = \, 2.70$, $|\varepsilon_{e\tau}|\, = \, 0.65$, and $\phi_{e\tau}\, = \,-170^{\circ}$ for the Best Reconstruction case. The best fit values of $\theta_{13}$, $\theta_{23}$, and $\Delta m_{31}^2$ have also been chosen from the simulation for both the standard and the NSI scenarios. Lower left (right) panel represents the absolute difference in the upper left (right) appearance events between the standard and the NSI scenarios divided by the statistical uncertainty in each bin. The color coding is the same as in the upper panels.}
\label{fig:spec_ee_etau}
\end{figure}

To be more explicit, we compare in Fig.~\ref{fig:spec_ee_etau} the appearance spectra between the standard case (solid lines) and the NSI case (dashed lines) corresponding to a point in parameter space which is ruled out by the Best Reconstruction but not by the TDR configuration (see the black star in Fig.~\ref{fig:2Dproj1} upper left panel).  Upper left (right) panel in Fig.~\ref{fig:spec_ee_etau} presents the $\nu_{\mu} \rightarrow \nu_e$ ($\bar{\nu}_{\mu} \rightarrow \bar{\nu}_e$) appearance event spectra as a function of reconstructed energy. Green and blue lines correspond to the TDR and Best Reconstruction configurations respectively. Solid lines represent the standard three-neutrino spectra with $\delta_{\rm CP}\,=\, -90^{\circ}$ and dashed lines represent the spectra with NSI in the $e-e$ and $e-\tau$ sector simultaneously. In case of NSI, we have taken the best fit parameters corresponding to the black star in the upper left panel of Fig.~\ref{fig:2Dproj1}, i.e., $\delta_{\rm CP}\,=\, -22^{\circ}$, $\varepsilon_{ee}\, = \, 2.70$, $|\varepsilon_{e\tau}|\, = \, 0.65$, and $\phi_{e\tau}\, = \,-177^{\circ}$ for the TDR, and $\delta_{\rm CP}\,=\, -23^{\circ}$, $\varepsilon_{ee}\, = \, 2.70$, $|\varepsilon_{e\tau}|\, = \, 0.65$, and $\phi_{e\tau}\, = \,-170^{\circ}$ for the Best Reconstruction case respectively. The best fit values of $\theta_{13}$, $\theta_{23}$, and $\Delta m_{31}^2$ have also been chosen from the simulation for both the standard and the NSI scenarios respectively. In the lower panels we represent the absolute difference $\Delta$ in the appearance events between the standard and the NSI scenarios divided by the statistical uncertainty in each bin. The color coding is the same as in the upper panels.
Here, the effect of the improved energy resolution is very clear: both first and second  maxima become statistically more powerful in rejecting this NSI hypothesis.
Although the second maximum has less events and less bins, the relative change in the number of events is significantly larger than in the first maximum.
This evidences that both oscillation maxima can contribute comparably to a better sensitivity, if they are resolved more precisely.

\section{Conclusions} 
\label{sec:con}
In this work we have investigated the impact of improved energy resolution on DUNE sensitivity to new physics. 
As an illustrative example, we have shown how a better energy resolution could lead to improved constraints on neutrino non-standard interactions with matter.
We have found that a better energy resolution would help to disentangle degeneracies among effects arising from standard three-neutrino oscillations, such as the usual $CP$ violation, and those stemming from NSIs.
In fact, by studying in more detail certain benchmark points, we have identified that the improved sensitivity to NSIs comes not only from the first oscillation peak, but the second maximum also contributes significantly to the experimental sensitivity as long as it can be appropriately resolved.
Therefore, a better neutrino energy determination allows the experiment to further leverage its broad-band beam in BSM physics searches.

More specifically, we have found that potentially significant sensitivity improvements could be achieved for $\varepsilon_{ee}$, $\varepsilon_{\mu\mu}$, $\varepsilon_{\tau\tau}$, and $\varepsilon_{e\tau}$, particularly when the NSI $CP$ violating phase is considered for $\varepsilon_{e\tau}$.
We have also found that certain degeneracies could be better resolved with an improved energy resolution.
For any combination of two of the aforementioned NSI parameters, we find a significant reduction in the allowed regions, when going from the CDR case to the TDR and especially to the Best Reconstruction scenario (see Fig.~\ref{fig:2Dproj1}).
Finally, as a by-product of our analysis, we have also shown how the $CP$ violation sensitivity and the $CP$ phase precision would improve with the neutrino energy resolution (see Fig.~\ref{fig:delcp_precision}). 
We hope our analysis will further motivate the pursuit of an enhanced neutrino energy reconstruction in liquid argon time projection chambers.

\subsubsection*{Acknowledgments}
We would like to thank Shirley Li for useful discussions on the DUNE energy resolution, Chris Marshall and Elizabeth Worcester for discussions and clarifications regarding the GLoBES simulation for DUNE TDR, and Sanjib Kumar Agarwalla for some earlier discussions on NSI at DUNE. Fermilab is operated by the Fermi Research Alliance, LLC under contract No. DE-AC02-07CH11359 with the United States Department of Energy. This project has received support from the European Union’s Horizon 2020 research and innovation programme under the Marie Skłodowska-Curie grant agreement No 860881-HIDDeN.
The work of BD is supported in part by the US Department of Energy under Grant No. DE-SC0017987, by the Neutrino Theory Network Program, and by a Fermilab Intensity Frontier Fellowship. BD and SSC would like to thank the Fermilab Theory Group for local hospitality during a summer visit in 2019, where this project was initiated.

\bibliographystyle{JHEP}
\bibliography{NSI-References}

\end{document}